\documentclass[twocolumn]{aastex631}
\usepackage[T1]{fontenc}

\newcommand{\Hcal}{\mathcal{H}}
\newcommand{\grad}{\boldsymbol{\nabla}} 
\newcommand{\di}{\partial} 
\newcommand{\e}{{\rm e}} 
\newcommand{\apgt}{\ {\raise-.5ex\hbox{$\buildrel>\over\sim$}}\ }
\newcommand{\aplt}{\ {\raise-.5ex\hbox{$\buildrel<\over\sim$}}\ }

\definecolor{Red}{rgb}{0.743,0,0}
\definecolor{Blue}{rgb}{0.25,.41,.88}
\definecolor{Green}{rgb}{0,0.5,0}

\def\Fbold{{\bf F}} 
\def\Jbold{{\bf J}}
\def\gbold{{\bf g}}

\def\fourt{{\textstyle{1\over4}}}

\def\twonine{{\textstyle{2\over9}}}
\def\grad{{\nabla}}

\def\deg{\ifmmode {^\circ}\else {$^\circ$}\fi}
\def\degree{\ifmmode {^\circ}\else {$^\circ$}\fi}
\def\mum{\ifmmode {\rm \,\mu {\rm m}}\else $\rm \,\mu {\rm m}$\fi}
\def\arcsec{\ifmmode ^{\prime \prime}\else $^{\prime \prime}$\fi}

\def\inch{\ifmmode ^{\prime \prime}\else $^{\prime \prime}$\fi}
\def\msunyr{\ifmmode {M_{\odot}~{\rm yr^{-1}}}\else $M_{\odot}~{\rm yr^{-1}}$\fi}
\def\msun{\ifmmode {M_{\odot}}\else $M_{\odot}$\fi}
\def\rsun{\ifmmode {R_{\odot}}\else $R_{\odot}$\fi}
\def\lsun{\ifmmode {L_{\odot}}\else $L_{\odot}$\fi}
\def\mstar{\ifmmode {M_{\star}}\else $M_{\star}$\fi}
\def\rstar{\ifmmode {R_{\star}}\else $R_{\star}$\fi}
\def\tstar{\ifmmode {T_{\star}}\else $T_{\star}$\fi}
\def\lstar{\ifmmode {L_{\star}}\else $L_{\star}$\fi}
\def\md{\ifmmode {M_d}\else $M_d$\fi}
\def\ld{\ifmmode {L_d}\else $L_d$\fi}
\def\ad{\ifmmode A_d\else $A_d$\fi}
\def\ldlstar{\ifmmode L_d / L_\star\else $L_d / L_{\star}$\fi}
\def\rearth{\ifmmode {\rm R_{\oplus}}\else $\rm R_{\oplus}$\fi}
\def\mearth{\ifmmode {\rm M_{\oplus}}\else $\rm M_{\oplus}$\fi}
\def\qdstar{\ifmmode Q_D^\star\else $Q_D^\star$\fi}
\def\kms{\ifmmode {\rm km~s^{-1}}\else $\rm km~s^{-1}$\fi}
\def\ms{\ifmmode {\rm m~s^{-1}}\else $\rm m~s^{-1}$\fi}
\def\mesc{\ifmmode m_{esc}\else $m_{esc}$\fi}
\def\rmin{\ifmmode r_{min}\else $r_{min}$\fi}
\def\rmax{\ifmmode r_{max}\else $r_{max}$\fi}
\def\mmin{\ifmmode m_{min}\else $m_{min}$\fi}
\def\mmax{\ifmmode m_{max}\else $m_{max}$\fi}
\def\rmind{\ifmmode r_{min,d}\else $r_{min,d}$\fi}
\def\rmaxd{\ifmmode r_{max,d}\else $r_{max,d}$\fi}
\def\mmaxd{\ifmmode m_{max,d}\else $m_{max,d}$\fi}
\def\vrad{\ifmmode v_{rad}\else $v_{rad}$\fi}
\def\qz{\ifmmode q_{0}\else $q_{0}$\fi}
\def\qi{\ifmmode q_{i}\else $q_{i}$\fi}
\def\ql{\ifmmode q_{l}\else $q_{l}$\fi}
\def\qs{\ifmmode q_{s}\else $q_{s}$\fi}
\def\rbrk{\ifmmode r_{brk}\else $r_{brk}$\fi}
\def\rdamp{\ifmmode r_{damp}\else $r_{damp}$\fi}
\def\rin{\ifmmode r_{in}\else $r_{in}$\fi}
\def\rout{\ifmmode r_{out}\else $r_{out}$\fi}
\def\tin{\ifmmode t_{in}\else $t_{in}$\fi}
\def\tout{\ifmmode t_{out}\else $t_{out}$\fi}
\def\ain{\ifmmode a_{in}\else $a_{in}$\fi}
\def\aout{\ifmmode a_{out}\else $a_{out}$\fi}
\def\r0{\ifmmode R_{0}\else $R_{0}$\fi}
\def\m0{\ifmmode m_{0}\else $m_{0}$\fi}
\def\M0{\ifmmode M_{0}\else $M_{0}$\fi}
\def\xm{\ifmmode x_{m}\else $x_{m}$\fi}
\def\sigz{\ifmmode \Sigma_0\else $\Sigma_0$\fi}
\def\gyr{\ifmmode {\rm g~yr^{-1}}\else ${\rm g~yr^{-1}}$\fi}
\def\cms{\ifmmode {\rm cm~s^{-1}}\else ${\rm cm~s^{-1}}$\fi}
\def\gcms{\ifmmode {\rm g~cm^{-2}}\else $\rm g~cm^{-2}$\fi}
\def\gcmss{\ifmmode {\rm g~cm^{-2}~s^{-1}}\else $\rm g~cm^{-2}~s^{-1}$\fi}
\def\gcmc{\ifmmode {\rm g~cm^{-3}}\else $\rm g~cm^{-3}$\fi}
\def\dcm2{\ifmmode {\rm dyn~cm^{-2}}\else $\rm dyn~cm^{-2}$\fi}
\def\ecsk{\ifmmode {\rm erg~cm^{-1}~s^{-1}~K^{-1}}\else $\rm erg~cm^{-1}~s^{-1}~K^{-1}$\fi}
\def\cm2{\ifmmode {\rm cm^{-2}}\else $\rm cm^{-2}$\fi}
\def\atilin{\ifmmode {\tilde{a}_{in}}\else $\tilde{a}_{in}$\fi}
\def\atilout{\ifmmode {\tilde{a}_{out}}\else $\tilde{a}_{out}$\fi}
\def\atil{\ifmmode {\tilde{a}}\else $\tilde{a}$\fi}
\def\ttil{\ifmmode {\tilde{t}}\else $\tilde{t}$\fi}
\def\sqrttt{\ifmmode {\tilde{t}^{1/2}}\else $\tilde{t}^{1/2}$\fi}

\def\h2o{H$_2$O}
\def\sio2{SiO$_2$}
\def\ch4{CH$_4$}
\def\h2{H$_2$}
\def \ms{m\,s$^{-1}$\,}
\def \kms{km\,s$^{-1}$}
\def \msun{M$_{\odot}$}
\def \rsun{R$_{\odot}$}
\def \lsun{L$_{\odot}$}

\def \mearth{M$_{\oplus}~$}

\shorttitle{Effect of Accretion Rate on Planetary Structure}
\shortauthors{Lozovsky et al.}

\graphicspath{{./}{figures/}}

\begin{document}

\title{The Effect of Accretion Rate and Composition on the Structure of Ice-rich Super-Earths}

\author[0000-0001-5451-3221]{Michael Lozovsky}
\affiliation{Department of Geosciences,\\ Tel-Aviv University, Tel-Aviv,\\ Israel}

\author[ 0000-0002-6317-4839 ]{ Dina Prialnik }
\affiliation{Department of Geosciences,\\ Tel-Aviv University, Tel-Aviv,\\ Israel}

\author[ 0000-0003-4801-8691 ]{Morris Podolak}
\affiliation{Department of Geosciences,\\ Tel-Aviv University, Tel-Aviv,\\ Israel}

\begin{abstract}

It is reasonable to assume that the structure of a planet and the interior distribution of its components are determined by its formation history. We thus follow the growth of a planet from a small embryo through its subsequent evolution. We estimate the accretion rate range based on a protoplanetary disk model at a large enough distance from the central star, for water ice to be a major component. We assume the accreted material to be a mixture of silicate rock and ice, { {with no H-He envelope, as the accretion timescale is much longer than the time required for the nebular gas to dissipate.}}  We adopt a thermal evolution model that includes accretional heating, radioactive energy release, and separation of ice and rock. Taking the Safronov parameter and the ice-to-rock ratio as free parameters, we compute growth and evolutionary sequences for different parameter combinations, for 4.6~Gyr.

We find the final structure to depend significantly on both parameters. Low initial ice to rock ratios and high accretion rates, each resulting in increased heating rate, lead to the formation of extended rocky cores, while the opposite conditions leave the composition almost unchanged and result in relatively low internal temperatures. When rocky cores form, the ice-rich outer mantles still contain rock mixed with the ice. We find that a considerable fraction of the ice evaporates upon accretion, depending on parameters, { {and assume it is lost,}} thus the final surface composition and bulk density of the planet do not necessarily reflect the protoplanetary disk composition.




\end{abstract}


\section{Introduction} \label{sec:intro}

We now have observational data on more than 5,000 exoplanets. These planets show very large diversity in their measured properties, such as mass and radius. While the majority of these planets have a Jupiter mass or more, super-Earth-sized planets are also very abundant \citep[e.g.][]{Fulton2017,Fressin2013,Batalha2013,Bean2021}. Gas giants are expected to have large gaseous envelopes (as their name implies), and it is not surprising that in a given mass range the radii of these bodies show a large variation \citep[e.g.][]{Laughlin2018,Hou2022}.  Smaller bodies, however, with masses of only a few $M_{\oplus}$, are less likely to accumulate large gaseous envelopes, and we would expect to see a correspondingly smaller variation in their radii. Yet, observed radii in this mass range can vary widely \citep[see, e.g.][Figure 3]{zeng2014}.  

Many factors contribute to this variation.  Obvious candidates are the presence or absence of a significant gaseous envelope, differences in internal composition, and surface temperature as fixed by the distance from the central star and temperature of that star.  However, there are also more subtle factors that are less commonly considered.  \cite{zeng2014} have pointed out that the slow cooling of a water-rich planet will cause noticeable changes in radius as a function of time.  Thus it is of interest to understand how the initial temperature and composition profiles of a planet might be dependent on the circumstances of its formation.  

The connection between planetary origin and internal structure, is nontrivial, and different formation mechanisms and birth environments can lead to a large range of compositions and internal structures \citep[e.g.][and references within]{Helled2014,Lozovsky2017,Valletta2020}. Even the host stellar type might influence the planetary structure and composition \citep{Lozovsky2021,Silva2016,Adibekyan2021}. While there are numerous observations of protoplanetary disks \citep[e.g.][]{Andrews2020}, and various aspects of their evolution have been modeled theoretically, the properties of such protoplanetary disks are still not well constrained.  Indeed, we still don't have a complete picture of the evolution of even our own proto-solar nebula \citep[e.g.][]{Villenave2021}.

In this paper we address one very specific aspect of the connection between formation processes and the resultant planetary structure:  For a planet formed by accretion of water and rock, how do the temperature and composition profiles depend on the accretion rate and planetesimal composition?  In view of the uncertainties involved, we consider a simple model of the protoplanetary disk and parameterize the associated accretion process.  We limit ourselves to the case where the planet is formed far from the central star (40\,au) and assume a simple model of planetary accretion.  In this way the amount of solid material for planet formation is low enough so that the planet does not reach the stage of rapid gas accretion before the disk dissipates \citep{Pollack1996,Alibert2005,Frelikh2017}. In addition, the protoplanetary embryo remains small enough while the gas is in place, so that migration is not an issue \citep{Armitage2020,Ogihara2022}.  In the case of pebble accretion things will happen more quickly and we address this in section\,\ref{discussion}.

In Section\,\ref{sec:accretion} we describe the nebular model and the parameterization of the accretion process, in Section \ref{sec:ThermalEvol} we describe the thermal evolution model used to evolve the planet, in Section\,\ref{sec:results} we present our results, and in Section \ref{discussion} we present our conclusions. 

\section{Accretion model} 
\label{sec:accretion}

\subsection{Protoplanetary Disk}

We assume a sun-like star (1 M$_\odot$) surrounded by a protoplanetary disk with a surface density of gas given by \citep{hayashi1981}
\begin{equation}
	\Sigma_g(a)=\Sigma_0 a^{-3/2} 
\end{equation}
where $a$ is the radial distance in au, and $\Sigma_0$ is the value of $\Sigma_g$ at $a=$1\,au.  If we assume that the disk extends to 50\,au and has a mass of $M_d=0.02M_{\odot}$, then we have 
\begin{equation}\label{diskmi}
	M_d=2\pi\int_{a_1}^{a_2}r\Sigma_g(r)dr
\end{equation}
where $a_1$ and $a_2$ are the inner and outer radii of the disk, respectively.  Setting $M_d=4\times 10^{31}$\,g and taking $a_1=0.01$\,au and $a_2=50$\,au in Equation (\ref{diskmi}) gives $\Sigma_0=2\times 10^3$\,\gcms.  The value of $\Sigma_0$ is insensitive to the choice of $a_1$ so long as $a_1\ll a_2$.  

We consider the region at $\sim$ 40\,au, far from the influence of other massive bodies which might be forming, and at a low enough temperature so that almost all the material other than hydrogen and helium is in the form of solids.  We further assume that the mass fraction of these solids is 0.02, in accordance with solar composition \citep{Lodders2010}.  This gives a surface density of solids, $\Sigma_s\approx 0.16$\,\gcms.  In order to parameterize the growth rate in a simple, yet physically reasonable manner, we assume that the proto-planetary embryo grows by accreting the material in an annulus centered at 40\,au and having a width of $\Delta a$ on either side, where $\Delta a$ is related to the Hill's sphere radius of the final body.  This 
\textit{isolation mass} is given by \cite{Lissauer1987} as
\begin{equation}\label{eq:miso}
	M_{iso}=1.95\times 10^{24}\left( Ba^2\Sigma_s\right) ^{3/2}
\end{equation}
where, again, $a$ is in au, $B\sim 3-4$ is a dimensionless constant, and all the rest of the parameters are in cgs units.  Taking $B=3$ gives an isolation mass at 40\,au of $4.2\times 10^{28}$\,g or about $7\,M_{\oplus}$.

\subsection{Mass Accretion Rate}

If $R$ is the body's radius, $\rho_s$ is the space density of solids in the region, and $v$ is the random velocity of the planetesimals with respect to the growing embryo, the rate of growth is given by \citep{Lissauer1987}
\begin{equation}\label{saf1}
	\dot{M}=\frac{dM}{dt}=\pi R^2\rho_s v \left[ 1+\left(\frac{v_e}{v}\right)^2\right]
\end{equation}
where $v_e$ is the escape velocity from the embryo and the term in brackets is the gravitational enhancement factor \citep[see][]{safronov1972}.  This can be rewritten in terms of the surface density as \citep{Lissauer1987}

\begin{equation}\label{eq:saf2}
	\dot{M}(t)=\frac{\sqrt{3}\pi}{2} R(t)^2\Sigma_s(t) \Omega \left[ 1+2\Theta\right]
\end{equation}
where $\Theta=v_e^2/2v^2$ is the so-called \textit{Safronov parameter}, and $\Omega$ is the Kepler frequency.  \cite{Lissauer1987}, based on a generalization of the results of \citet{wetherill1985}, gives 
\begin{equation}\label{eq:safaprox}
\Theta\approx 400a\left(\frac{\bar{\rho}}{4.04}\right)^{1/3}
\end{equation}
where $\bar{\rho}$ is the mean density of the planetary embryo in cgs, and $a$ is again in au.  If we ignore the very weak dependence on $\bar{\rho}$, we find that at 40\,au the expected value is $\Theta=16,000$. We assume that the total mass (planetesimals + growing embryo) in this annulus remains constant with time.  In what follows, we allow $\Theta$ to vary around its fiducial value in order to explore how different accretion rates affect the temperature and composition profiles for a given mass.  It should be noted that alternative formation mechanisms, such as pebble accretion, will yield a very different $\dot{M}$ profile \citep[e.g.][]{Bitsch2015,Andama2021,Ormel2021}, and we address this issue in the discussion section. We define a normalized Safronov parameter for this study $\hat{\theta}=\frac{\Theta}{a} $,  {than a is in au units}.

\subsection{Composition}

As the planet grows in a region of $\sim$ 40 au, we assume that the planetesimals are composed of rock (SiO$_2$) and water ice (H$_2$O). We do not take into account H-He gas, since the prescription we have chosen for $\dot{M}$ assures that the nebular gas will dissipate while the mass of the protoplanetary embryo is still very small. For simplicity, heavier elements such as metals are not considered, as their contribution to the mass-radius relation is weak \citep{Vazan2018}.  Thus, the rock is assumed to be one single substance, and we ignore possible chemical reactions or differentiation within the rock, such as separation of iron from silicates. The ice is also taken as one substance, thus phase transitions between various forms of solid ice are ignored and only melting/freezing is taken into account.

 The ice mass fraction we investigate ranges from 0.3 to 0.7 out of the total mass of the material available for accretion in the disk. After accretion, the ice and rock distribution depends on the tendency of those two substances  to mix or separate in different regimes. A separation between ice and rock may lead to formation of layers, based on their densities \citep{Stevenson2013,Vazan2020}.  Since the phase separation between rock and ice is temperature and pressure dependent \citep[e.g.][]{Dorn2021,Vazan2020}, we investigate how different mass fractions of ice influence the formation of different cores. { {The composition of the accreted material is the same throughout the accretion process, but the radioactive content corresponds to the accretion time, that is, diminishes with time as in the planet itself.}}
 
\section{Thermal Evolution Model}
\label{sec:ThermalEvol}

We consider a spherical body that grows from an initial embryo of small mass compared to the final mass. The composition of both initial embryo and accreted material is a mixture of rock and ice. After $\hat\theta$, the ratio of ice to rock is the second free parameter of this study. The mass fraction of ice is denoted $X_i$ and that of rock, $X_r=1-X_i$. If melting occurs, a fraction of $X_i$ will be liquid; it is taken as a smoothed step function of the temperature around the pressure-dependent melting temperature $T_m(P)$, $X_\ell=X_i/[1+{\rm e}^{\beta(1-T/T_m)}]$ \citep[see][for further details]{Prialnik2008}. When the ice is heated and eventually melts, its viscosity $\eta(T,T_m)$ decreases and the rock settles toward the center, allowing ice-rock separation to occur. This is modeled by a downward flux of rocky material $\Jbold_r$ and an equal upward flux of ice/water $\Jbold_i$; we assume the relative movement to be symmetrical with respect to the local center of mass. Thus ice/water is emplaced by rock and thermal energy is exchanged in the process, since the specific energies of ice/water and rock differ. The relative velocity of rock and ice is taken to be
\begin{equation}
    v_r=\twonine(\rho_r-\rho_i)gr_p^2/\eta,
\end{equation}
(Stokes flow) where $\rho_r$ and $\rho_i$, are the densities of rock and ice, respectively, $g$ is the gravitational acceleration and $r_p$ is a typical radius of a rock particle. Since the movement is slow,
it is assumed that thermodynamic and hydrodynamic equilibrium are maintained, thus there is a unique local temperature and the structure evolves quasi-statically.

When a body grows in mass by accretion, it also gains thermal energy from the kinetic energy of the accreted material. { {Since the relative velocity of planetesimals entering the activity sphere of the accreting body is lower than the free-fall velocity, the rate of gain of accretion energy is only a fraction of $GM\dot M/R$ \citep{Bodenheimer1986, Boujibar2020}. But even then, not }} all the kinetic energy turns into heat; a large fraction goes into mechanical energy, such as excavating a crater or breaking rocks \citep{Ransford1982}. { {As working hypothesis, we}} assume the fraction that turns into heat to be $\fourt$ of the kinetic energy obtained if the accreted material free-falls from infinity \citep[c.f.][]{Bodenheimer1986, Pollack1986}. We further assume that the accretion energy is divided between a surface and a body source \citep{Squyres1988}. A fraction $f$ is taken to be deposited at the surface, while the remaining fraction $(1-f)$ is absorbed in the interior, the effect diminishing with depth. { {We adopt $f=0.4$, but the results are not very sensitive to the exact value.}}

We solve the coupled energy and mass conservation equations. For energy,
\begin{equation}
\frac{\di(\rho u)}{\di t} + \grad\cdot\Fbold + \grad\cdot(u_r\Jbold_r 
 + u_i\Jbold_i) = \dot Q
\label{eq:cons_e} 
\end{equation}
where $\rho u$ denotes the sum of weighted specific energies, $\Fbold$ is the heat flux, { {$\Fbold=-K\grad T$, $K$ denoting the thermal conductivity,}} and $\dot Q$ includes all energy sources: insolation, { {accretion, compression by gravitational forces, latent heat (which may be negative)}}, and radioactive decay energy by long-lived radioisotopes embedded in the rock. With the rate of growth of the planetary mass $\dot M(t)$ given by Equation \ref{saf1}, the surface boundary condition becomes
\begin{equation}
F(R)=\sigma T^4+Z(T)\Hcal_v-(1-A)\frac{L}{4\pi a^2}-\fourt{} f\frac{GM\dot M}{4\pi R^3},
\end{equation}
where $\sigma$ is the Stefan-Boltzmann constant, $A$ is the albedo, $L$ - the stellar luminosity, and $a$ - the distance from the star. If the surface temperature is sufficiently high, the ice may sublimate; $Z(T)$ is the sublimation rate determined by the saturated vapor pressure, and $\Hcal_v$ is the corresponding latent heat absorbed. For mass,
\begin{equation}
\frac{\di (\rho X_r)}{\di t} + \grad\cdot\Jbold_r = 0
\label{eq:cons_rhoi}
\end{equation}
\begin{equation}
\frac{\di (\rho X_i)}{\di t} + \grad\cdot\Jbold_i = 0.
\label{eq:cons_rhor}
\end{equation}
At the surface, $J_i=Z(T)$. Finally, the bulk density profile at any time $t$ is given by the hydrostatic equation,
\begin{equation}
    \grad P(\rho)-\rho\gbold=0,
    \label{eq:hydro}
\end{equation}
where $P(\rho)$ is the pressure given by the equation of state.

For a given pressure $P$ and ice mass fraction $X_i$, the bulk density weighted by the volume fractions of the components is given by
\begin{equation}
\label{eq:density1}
\frac{1}{\rho}= \frac{X_i}{\rho_i(P)}+\frac{1-X_i}{\rho_r(P)},
\end{equation}
where $\rho_i$ and $\rho_r$ are the ice and rock densities, respectively, for each of which we adopt the second-order Birch-Murnaghan (BM) approximation. The coefficients are chosen so as to achieve a close agreement with the tabulated equation of state (EOS) developed and used by e.g., \cite{vazan13, vazan15}, based on the quotidian EOS (QEOS) of \cite{more88}. Hence $\rho_i$ is obtained by solving
\begin{equation}
\label{eq:eosi}
P_{0,i}\left[\left(\frac{\rho_i(P)}{\rho_{0,i}}\right)^{7/3}-\left(\frac{\rho_i(P)}{\rho_{0,i}}\right)^{5/3}\right]-P =0,
\end{equation}
where $\rho_{0,i}=0.917$~g/cm$^3$, $P_{0,i}=2.07\times10^{11}$~dyn/cm$^2$, and similarly, $\rho_r$ is obtained by solving
\begin{equation}
P_{0,r}\left[\left(\frac{\rho_r(P)}{\rho_{0,r}}\right)^{7/3}-\left(\frac{\rho_r(P)}{\rho_{0,r}}\right)^{5/3}\right]-P=0,
\label{eq:eosd}
\end{equation}
where $\rho_{0,r}=2.5$~g/cm$^3$ and $P_{0,r}=1.28\times10^{12}$~dyn/cm$^2$.
In the range of interest for densities and temperatures, the effect of temperature on pressure according to the QEOS is very small. We have also estimated the Debye correction to the BM EOS and found it small. 
Therefore, we assume the pressure to be independent of temperature for the entire relevant temperature range.

The components of the volume energy source $\dot{Q}$ (energy per unit time per unit volume) in Eq.~(\ref{eq:cons_e}) are as follows:

The accretional heating source for the interior is
\begin{equation}
\dot{Q}_{\rm acc}=\fourt (1-f)\frac{3GM(t)\dot M(t)}{4\pi R^4(t)}h(z),
\end{equation}
where $h(z)$ is an exponentially decreasing function of the normalized depth $z$, such that $\int_0^1h(z)dz=1$. 

The radioactive energy source is
\begin{equation}
    \dot{Q}_{rad}= \rho_r\sum_j \tau_j^{-1}X_{0,j}H_j{\rm e}^{-t/\tau_j},
\end{equation}
where $X_{0,j}$ is the initial abundance (relative to rock) of the radionuclide, $H_j$ - the energy released per unit mass, and $\tau_j$ - the characteristic decay time. We consider $^{40}$K, $^{232}$Th, $^{238}$U and $^{235}$U.

The heat absorbed/released by melting/refreezing is given by  
\begin{equation}
 \dot{Q}_\ell=-\rho\frac{\di X_\ell}{\di t} H_\ell = -\rho
 X_i\frac{d}{dT}\left[\frac{1}{ \e^{\beta(1-T/T_m)} } \right]\frac{\di T}{\di t} H_\ell,
\end{equation}
where $H_\ell$ is the latent heat of melting.

As we are considering a spherically symmetric body, we choose the volume enclosed by a spherical surface of radius
$r$ ($0\ \le r\ \le R$), denoted by $V$ ($0\ \le V\ \le 4\pi R^3/3$) as the independent space variable.
Thus mass and energy fluxes are replaced by energy or mass crossing a spherical surface per unit time,
and
$$\grad \ \Longrightarrow \frac{\di}{\di V}.$$
In the numerical computations, where the equations are discretized in space, and the spatial grid is adapted
to the varying configuration of the model, we
consider a different, dimensionless space variable $x$, defined over a finite range $[c,s]$, where $c$ and $s$
are the system's boundaries (center and surface).
In this case $V(x)$ must be supplied as a monotonically increasing
solution of an equation, satisfying $V_c=0$ and $V_s=V_s(t)$ (moving boundary). 
The equation is so chosen as to ensure fine zoning where steep gradients (of temperature or other physical property) 
arise and has the general form
\begin{equation}
\frac {\di f(V)}{\di x} = {\rm constant}.
\end{equation}
Although the range of $x$ is fixed, the total volume $V(s)$ changes with time due to accretion of material, $\dot V(s)=\dot M/\rho(s)$, and consequently $V(x)$ changes at all $x$.
Since temporal derivatives are taken at constant $V$, whereas $V=V(x,t)$, the following transformation
is implemented in the difference scheme: 
\begin{equation}
\left(\frac{\di}{\di t}\right)_V = \left(\frac{\di}{\di t}\right)_x - \left(\frac{\di V}{\di t}\right)_x 
\left(\frac{\di}{\di V}\right)_t.
\end{equation} 
{ {The set of two-boundary time-dependent implicit difference equations is linearized and solved iteratively using the LINPACK library. Time steps are not limited because the difference scheme is implicit. Hydrostatic equilibrium is restored at each time step by solving numerically equation ~\ref{eq:hydro}. The corresponding change in gravitational potential energy, $E_g=-\int_0^{M(t)}\frac{Gmdm}{r}$ is calculated and 
we parametrize the distribution of this energy to obtain the local $\dot{Q}_{grav}$ for the timestep.
}}
The initial and physical parameters assumed for all cases are given in Table~\ref{tab:init}.

\begin{deluxetable*}{ll}
\tablenum{1}
\tablecaption{Initial and physical parameters}
\tablewidth{0pt}
\tablehead{
\colhead{Parameter} &  \colhead{Value}
}
\label{tab:init}
\startdata
Initial $^{40}$K abundance           &  $1.13\times10^{-6}$~ppm \\
Initial $^{232}$Th abundance      &  $5.52\times 10^{-8}$~ppm\\
Initial $^{235}$U abundance          &  $6.16\times 10^{-9}$~ppm\\
Initial $^{238}$U abundance &  $2.18\times 10^{-8}$~ppm\\
Albedo   &  0.5 \\
Ice specific energy                        &  $3.75\times 10^4 T^2 + 9.0\times 10^5 T$~erg~g$^{-1}$ \\
Water specific energy                   &  $4.187\times 10^7 T$~erg~g$^{-1}$ \\
Rock specific energy                    &   $1.3\times 10^7 T$~erg~g$^{-1}$ \\
Water thermal conductivity           & $5.5\times 10^4$~erg~cm$^{-1}$~s$^{-1}$~K$^{-1}$\\
Ice thermal conductivity           &  $5.67\times 10^7/T$~erg~cm$^{-1}$~s$^{-1}$~K$^{-1}$ \\ 
Rock thermal conductivity            &  $2\times 10^5$~erg~cm$^{-1}$~s$^{-1}$~K$^{-1}$ \\
Water dynamic viscosity               & $2.939\times10^{-4}\exp{\left(\frac{507.88}{T-149.3}\right)}$~dyn~cm$^{-2}$~s \\
                                                     &          [ $5.05\exp{(-5.71T/T_m)}$ for high $T_m$] \\
Latent heat of melting                   & $3.34\times 10^9$~erg~g$^{-1}$ \\
Latent heat of sublimation           &   $A_0- A_1T+A_2T^2-A_3T^3$~erg~g$^{-1}$, \\
 & $A_0=3.714\times10^{10},\ A_1=7.823\times10^7,$ \\
 &$A_2=1.761\times10^5,\ A_3=1.902\times10^2$ \\
\enddata 
{ {Note: The abundances of radioactive nuclides are those of chondritic meteorites scaled back 4.5~Gyr \citep{Anders1989}; for ice, the specific energy was derived by \citet{Klinger1980} based on data from \citet{Giauque1936} and the thermal conductivity is from \citet{Klinger1980} as well; for the thermal properties of rock, see \citet{Neumann2020} and also the recent paper by \citet{Bagheri2022}, based on new data from {\it{New Horizons}}; the viscosity of ice at high pressure is taken from \citet{Pigott2010}; the thermal conductivity of rock, from \citet{Seipold1998}, ignoring the relatively weak temperature dependence.}}
\end{deluxetable*}

%


\section{Results of evolutionary calculations}
\label{sec:results}

We start by describing the evolution of a prototype model in some detail. We then turn to discuss the effect of the two leading parameters on the evolution outcome.

\subsection{Evolutionary course of a typical case}
\label{ssec:typical}

\begin{figure}[h]
\centering
\includegraphics[width=0.475\columnwidth]{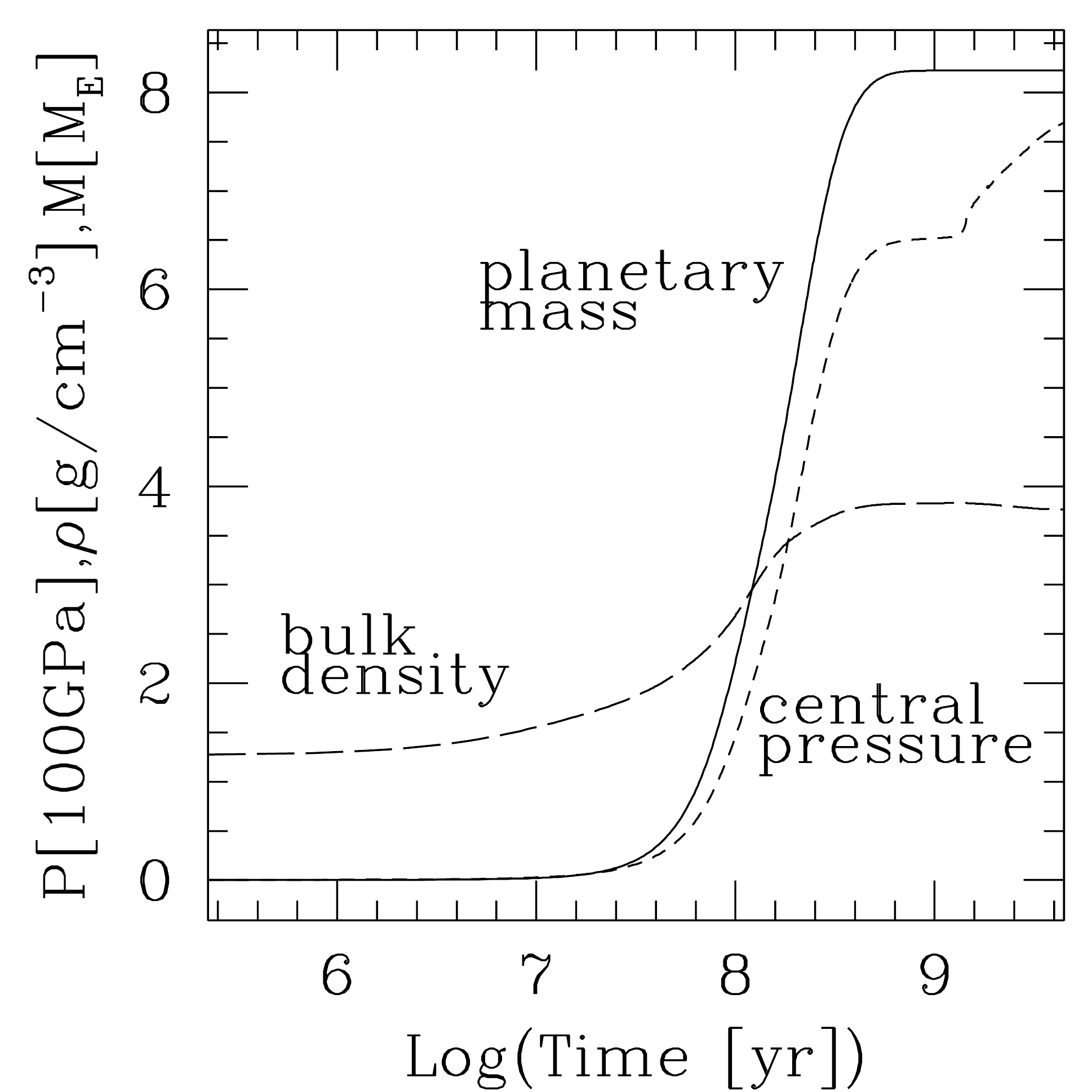}
\includegraphics[width=0.475\columnwidth]{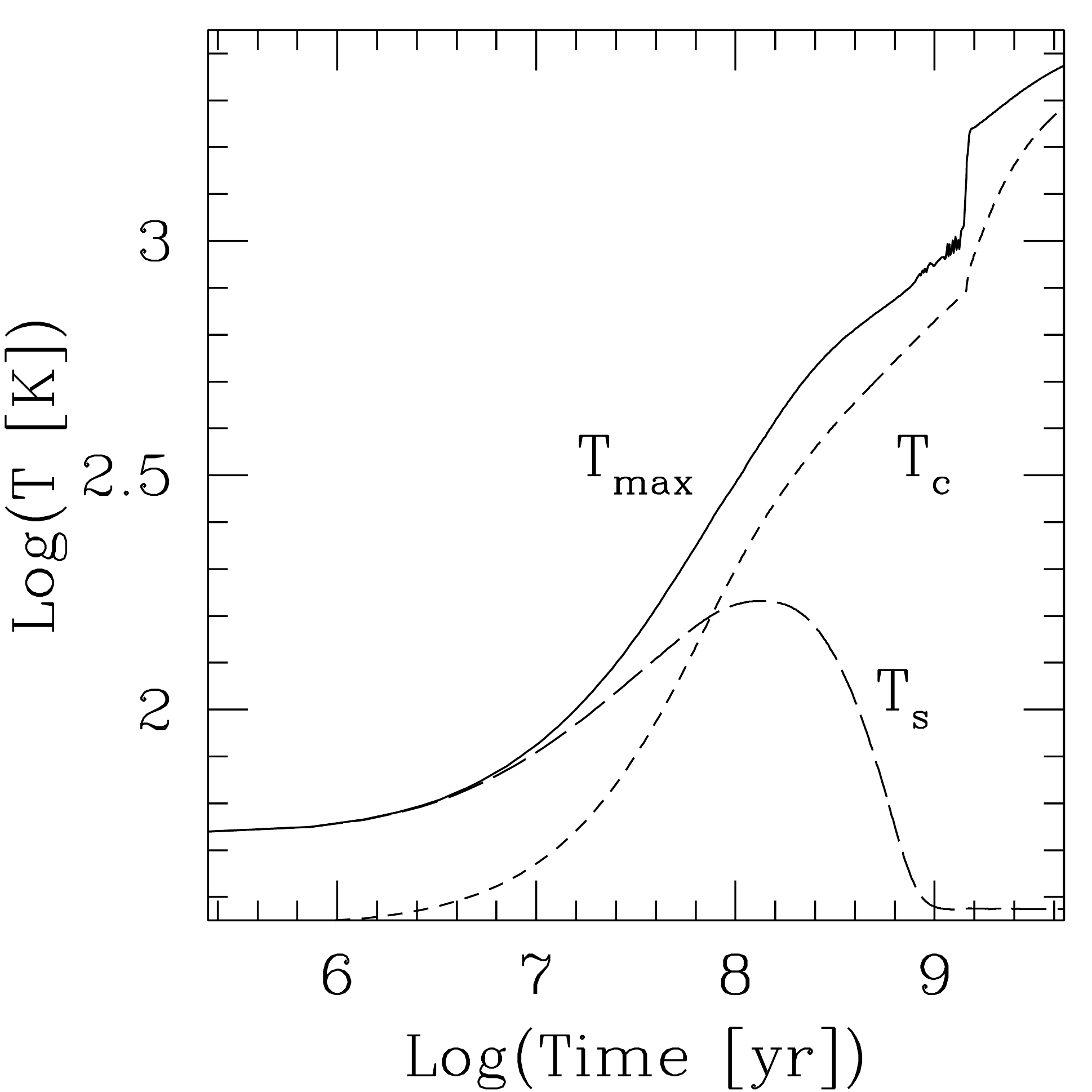}
\includegraphics[width=0.475\columnwidth]{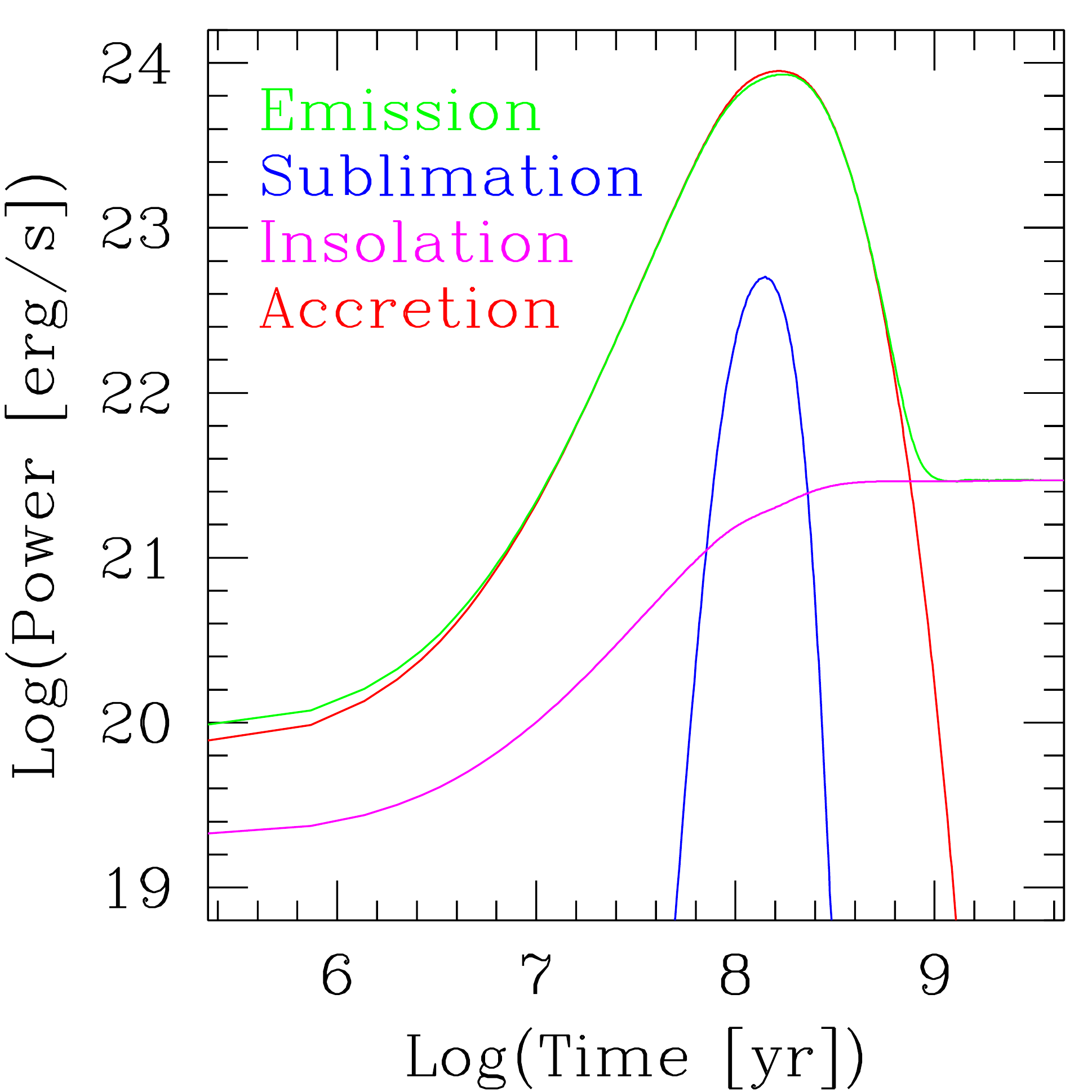}
\includegraphics[width=0.475\columnwidth]{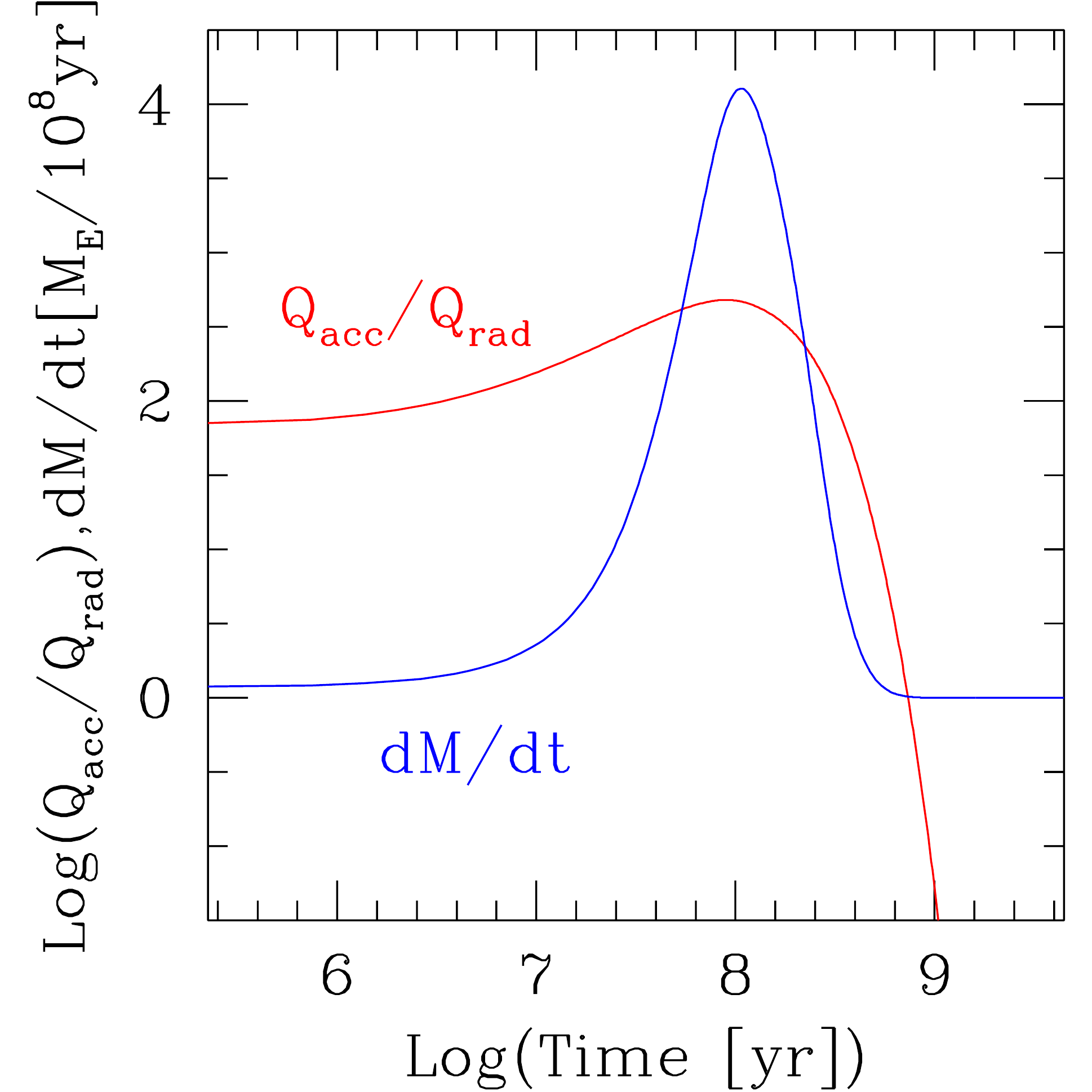}
\caption{Evolution of various properties for a model with $\hat\theta=400$ and $X_i=0.3$. Note that time is on a logarithmic scale, emphasizing changes in evolutionary timescales. {\textit{Upper left}} - the growing planetary mass, the bulk (average) density, and the central pressure. {\textit{Upper right}} - the maximal temperature, attained off-center, the central temperature and the surface temperature. {\textit{Lower left}} - energy per unit time supplied by accretion and insolation, absorbed by sublimation, and emitted by thermal radiation, where the small balance is conducted to/out of the interior. {\textit{Lower right}} - the ratio of energy released by accretional heating and by radioactive decay, and the accretion rate (for guidance).}
\label{fig:evol}
\end{figure}

We choose the case with normalized Safronov parameter $\hat\theta=400$ and a composition of 0.3 ice and 0.7 rock by mass. The initial embryo's mass is about 0.00015 of the final mass (roughly 0.1 lunar masses). The temperature is uniform and equal to the ambient temperature and the structure is in hydrostatic equilibrium. Accretion starts slowly, but picks up quickly, as shown in the lower right-panel of Figure~\ref{fig:evol}; the central pressure rises with the growing mass, as does the bulk density (see upper left-panel of Figure~\ref{fig:evol}). The accretion rate peaks at about $10^8$~yr, declines and ceases at $\sim 1$~Gyr. Accretional heating at the surface, which surpasses the solar energy by orders of magnitude, causes the surface temperature to rise up to $\sim$180~K. Although most of this heat is reradiated, a fraction is absorbed in ice sublimation, which is significant at this temperature (see lower left-panel of Figure~\ref{fig:evol}). When mass accretion is over, the planet settles into equilibrium, reradiating the stellar energy absorbed. 

\begin{figure}[h]
\centering
 \includegraphics[width=0.475\columnwidth]{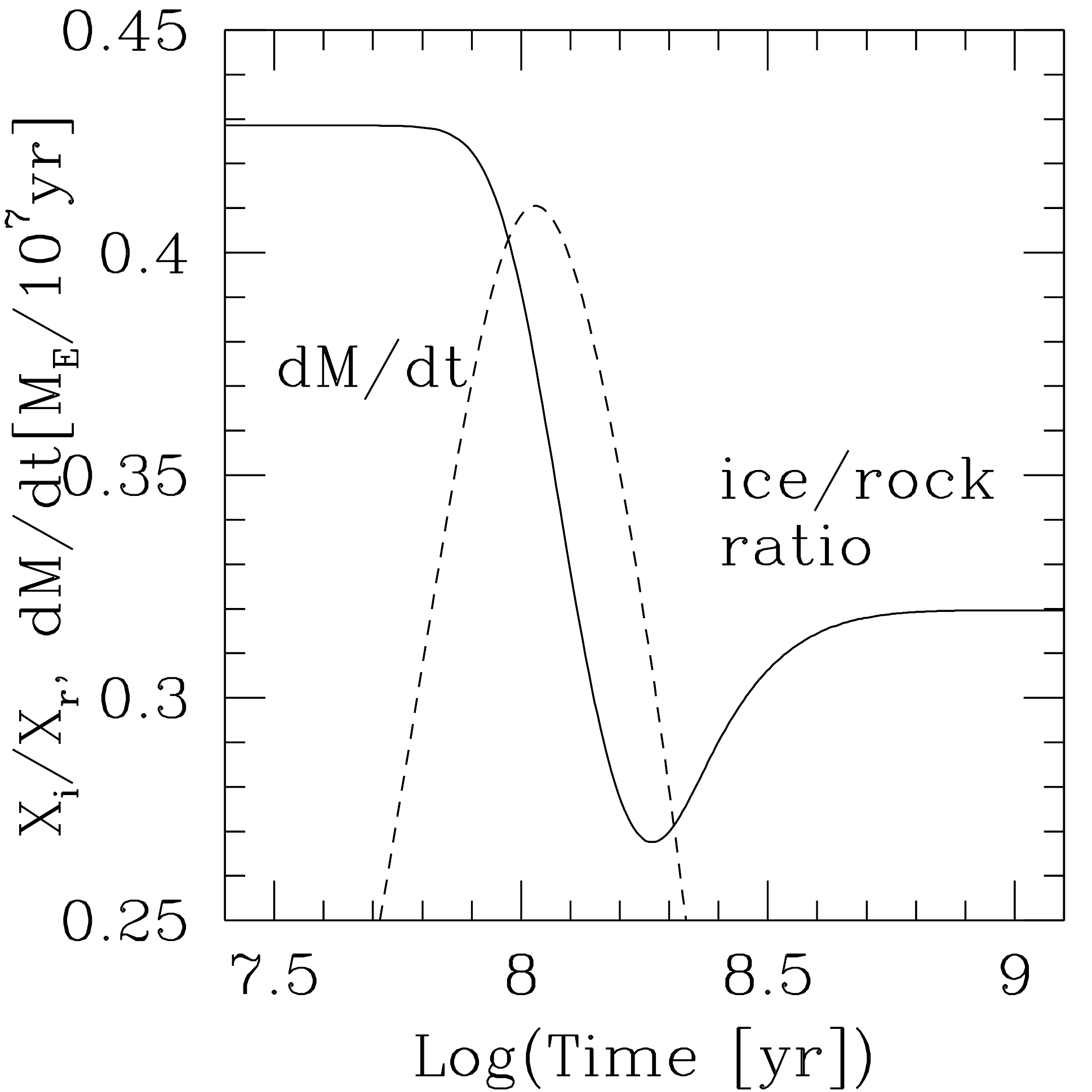}
\includegraphics[width=0.475\columnwidth]{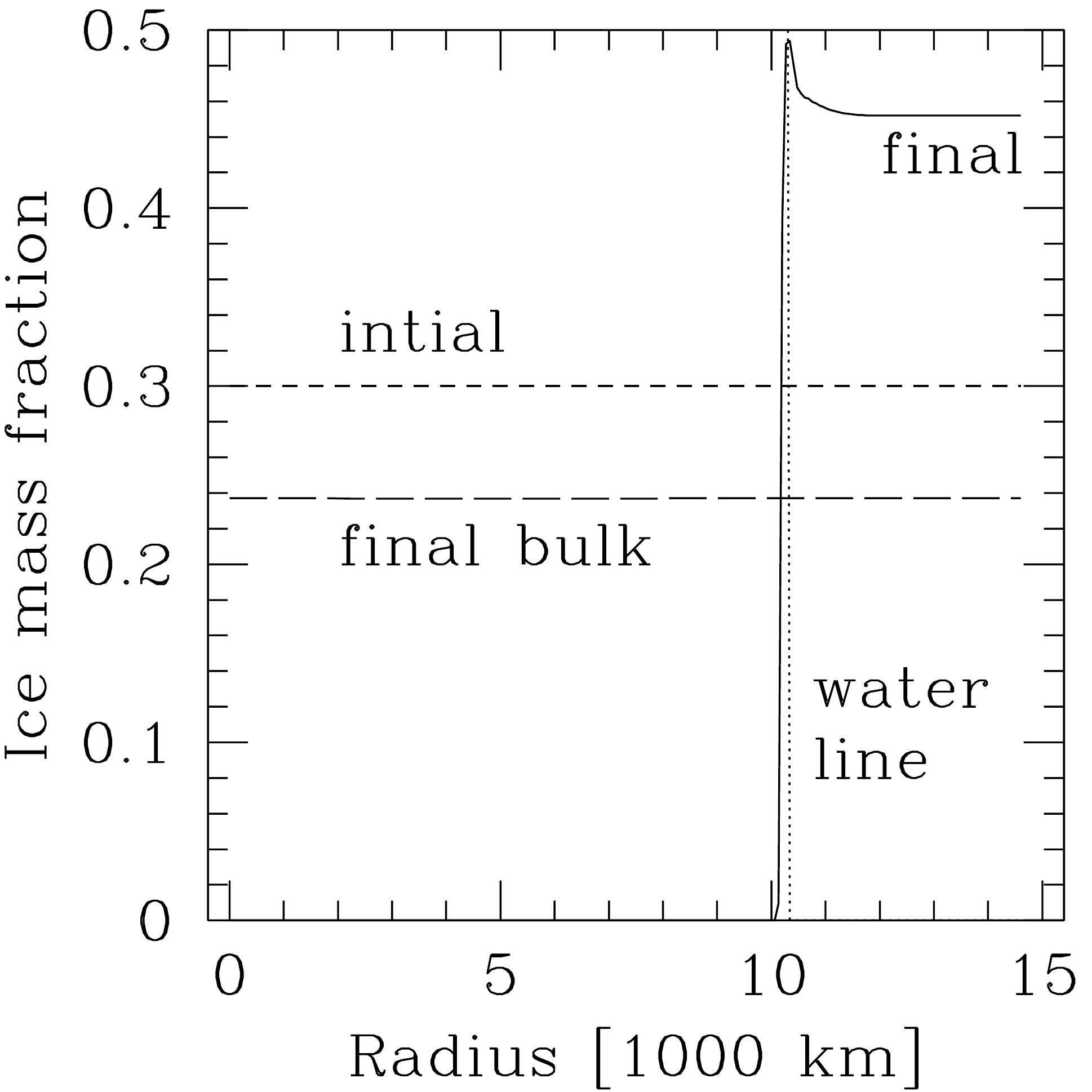}
\includegraphics[width=0.475\columnwidth]{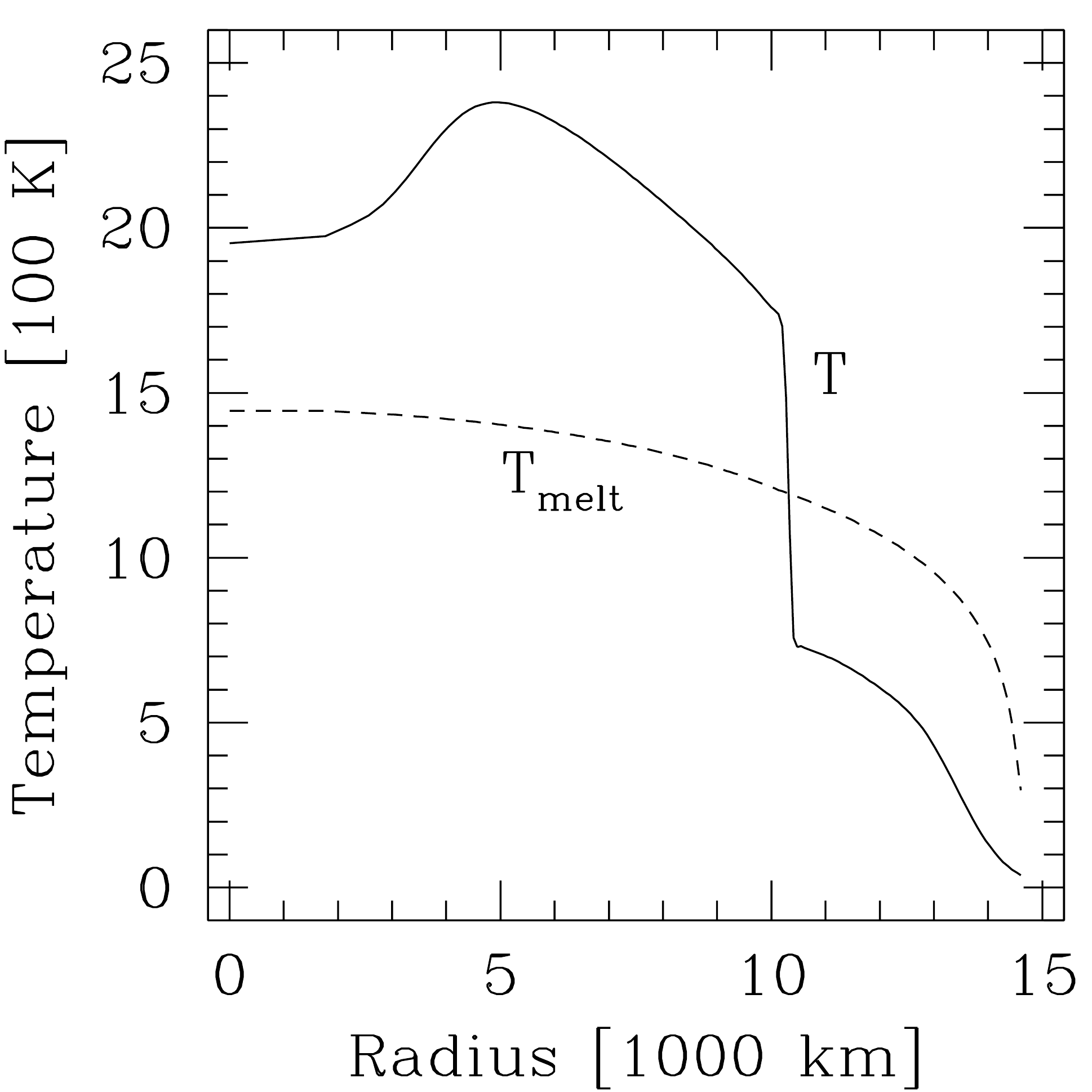} \includegraphics[width=0.475\columnwidth]{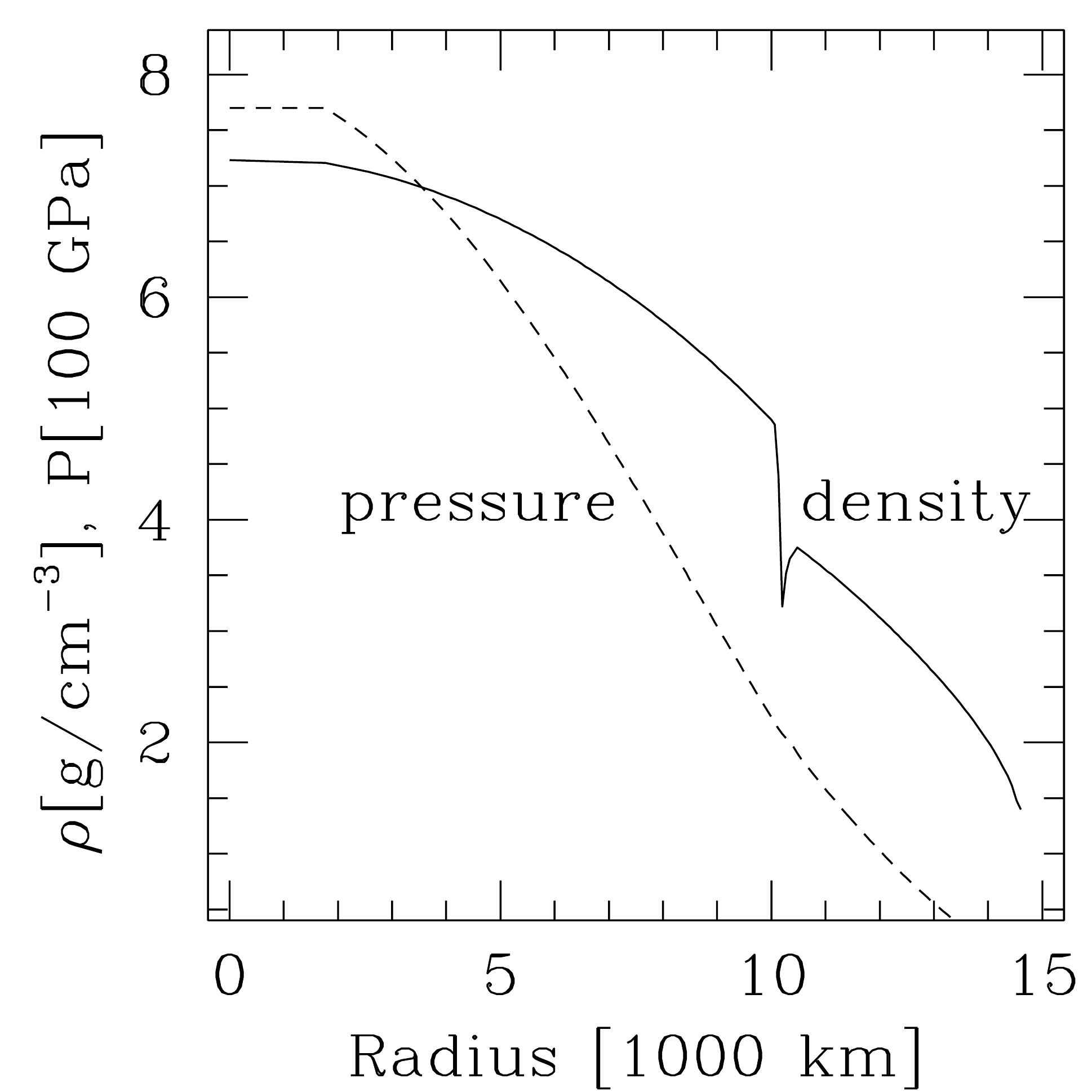}
    \caption{Characteristics of the model with $\hat\theta=400$ and $X_i=0.3$ (same as Figure~\ref{fig:evol}): {\textit{Upper-left}} - evolution of the bulk ice to rock ratio, reflecting the loss of ice at the peak of accretion. {\textit{Other panels}} - profiles of various characteristics throughout the planet at the end of evolution. In all profiles the core boundary can be clearly seen. The {\textit{water line}} marks the radial distance below which the temperature exceeds the melting temperature.}
    \label{fig:struct}
\end{figure}

Sublimation occurs both at the surface and in a porous subsurface layer, the vapor flowing to the surface and escaping. The loss of ice affects the bulk ice to rock ratio, which decreases during the phase of intense heating, but rises again when the surface temperature drops with the declining accretion rate, and most of the accreted ice is retained again. This effect is illustrated in the upper panels of Figure~\ref{fig:struct}, which show the variation of the bulk ice mass fraction with time (left) and its profile throughout the planet at the end of evolution (right). 

Accretional heating of the interior, mainly of the outer layers, surpasses heating by radioactive decay, therefore the maximal temperature is attained above the center. The heat wave penetrates inward and when it reaches the core and the ice reaches the melting temperature (above $10^3$~K at these pressures), the rock settles to the center and separation between ice and rock begins, forming a rocky core and an increasingly ice-rich mantle. This occurs at 1.8~Gyr and goes on to the present. The rocky core heats rapidly by radioactive decay and compression; the onset of core formation is marked by a rise in central (as well as in maximal) temperature and in central pressure, as seen in the upper panels of Figure~\ref{fig:evol}. Between the rocky core and the mantle there is a layer of liquid water mixed with rock, an internal ocean, about 150~km thick, but only a thin layer compared to the planet's size. This layer moves out as the core grows.

\begin{figure}[h]
\centering
 \includegraphics[width=1\columnwidth, trim={0.0cm 0.0cm 0.0cm 0.2cm },clip]{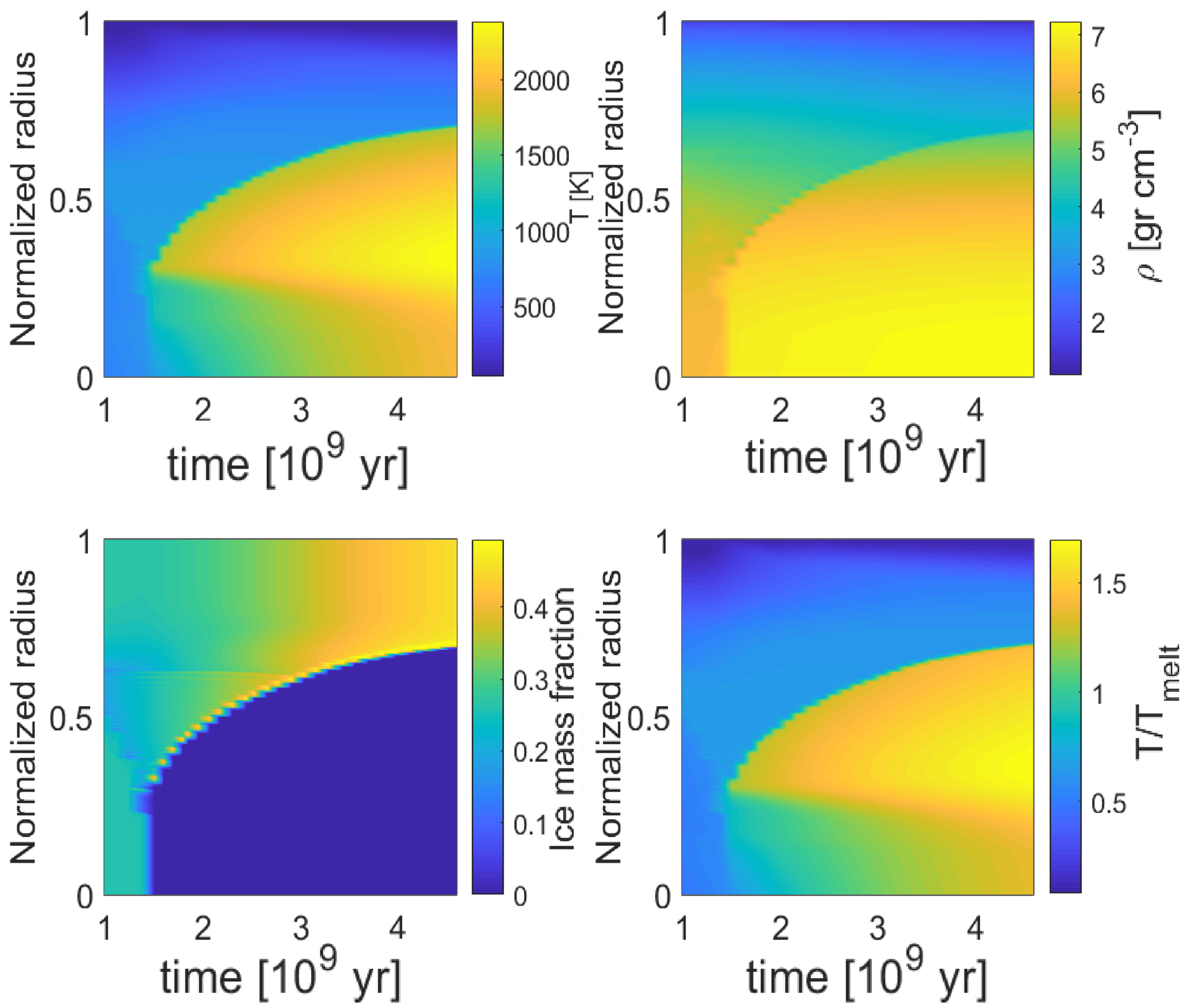}
    \caption{Evolution of planetary temperature, density, ice mass faction as a function of planetary radius and time. The time-frame is chosen to be such that the planetary radius is constant. $T_{melt}$ is melting temperature of water. The model assumed is $\hat{\theta}=400$ and ice mass fraction of 0.3.}
    \label{fig:surfplot}
\end{figure}

The continuous post-accretion evolution is shown in Figure~\ref{fig:surfplot}. We can see the advance of the surface heat wave toward the center and the marked effect of radioactive heating of the rock, once ice is separated from it. We also note that the core forms quickly at 2~Gyr and then expands slowly. At any given depth, the temperature rises with time.

The emerging structure after 4.5~Gyr of evolution is shown in Figure~\ref{fig:struct} by profiles of the main characteristics. The rocky core boundary is clearly seen by the sharp change in the temperature, density and ice mass fraction profiles (only the pressure is continuous). The core boundary coincides with the point where the temperature crosses the melting temperature curve. Although radioactive heating is stronger in the core, where the rock mass fraction is higher, the temperature peak occurs away from the center, even after the end of accretion. This is due to net heat transport outwards by the water, both because its heat capacity is higher than that of rock, and because it transports latent heat - absorbing heat by melting in deeper regions and releasing it by freezing in outer regions. We note that the ice content of the mantle is much higher than the original. The core extends to about 2/3 of the planet's radius.

We show in what follows that the extent of the core depends critically on the parameters that we investigate, to the point that it does not form at all. The history of ice accretion remains imprinted in the composition of the planet, where the ice content has a minimum at roughly half the mass at the end of the accretion phase. This memory is eventually erased when the core forms, but it remains present in models that do not form a core.

\subsection{Parameter Study}
\label{ssec:combinations}

\subsubsection{Evolution}

We now choose a baseline case that has a normalized Safronov parameter $\hat\theta_b=400$ and an ice mass fraction $X_{i,b}=0.5$ (ice to rock=1). To determine the effect of the accretion rate and correspondingly, of the accretion heating, we keep $X_{i,b}$ and vary $\hat\theta$ between 200 and 500. Similarly, to determine the effect of initial composition we keep $\hat\theta_b$ and vary the ice mass fraction between 0.3 and 0.7. All models start with a very small embryo ($\sim$0.00015M, where M is final mass) having the same composition as the accreted material.

\begin{figure*}
	\centering
	\includegraphics[width=0.9\textwidth, trim={0.5cm 1.4cm 0.0cm 0.0cm },clip]{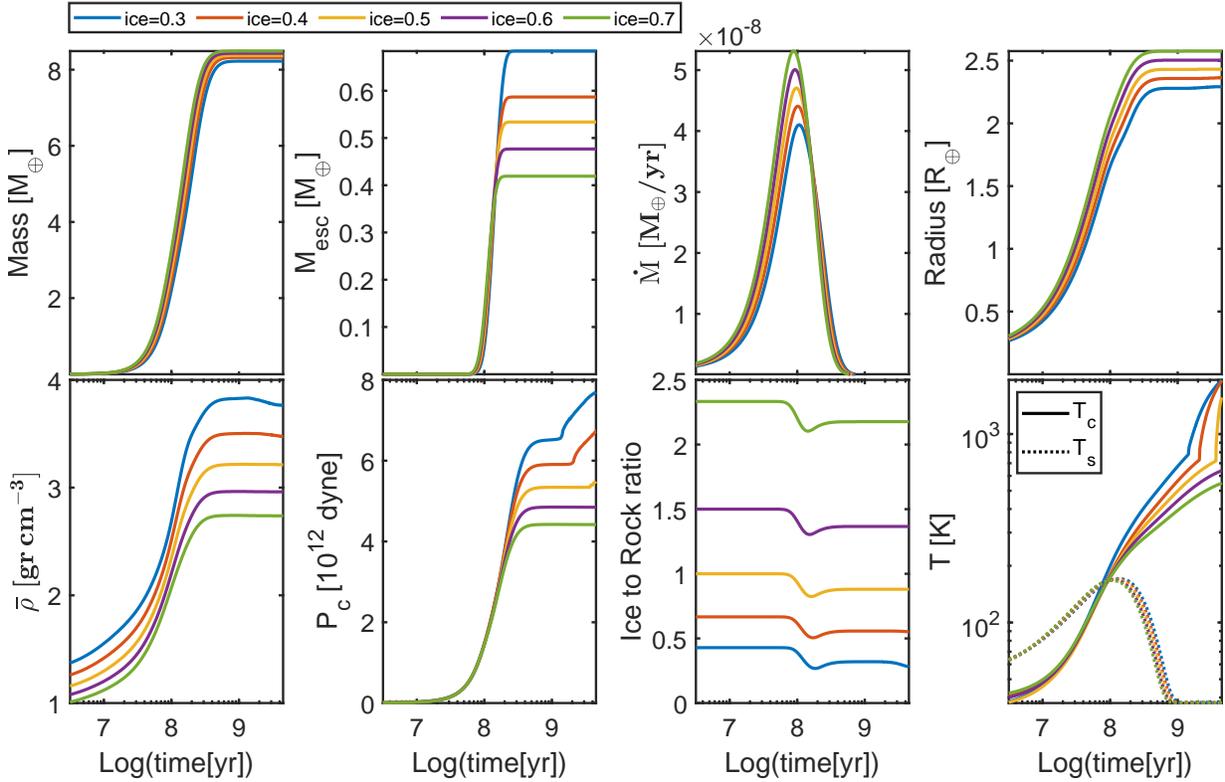}
		\caption{Various planetary properties versus time for different ice mass fraction. The models assume $\hat{\theta}=400$. Top row panels: the planetary mass $M(t)$; the mass of the escaping vapor $M_{\rm esc}(t)$; the mass accretion rate $\dot M(t)$; the planetary radius $R(t)$. Bottom row panels: the average (bulk) density $\bar\rho(t)$; the central pressure $P_c(t)$, the ice-to-rock mass ratio; the central and surface temperatures, $T_c(t)$ and $T_s(t)$ respectively.}
	\label{fig:varIce}
\end{figure*}

In Figure~\ref{fig:varIce}  we show evolution results for $\hat\theta_b$, and varying ice mass fraction.  In this set of models, the final mass is found to be almost independent of the planetesimal composition, while other properties depend on it strongly. As expected, the radius gradually increases with the ice mass fraction, while the bulk density $\bar\rho$ (fifth panel, Figure~\ref{fig:varIce}) and central pressure $P_c$ decrease. The mass accretion rate  $\dot{M}$, increases with the ice mass fraction, because the radius of the planet increases (see equation \ref{eq:saf2}). In addition, the peak becomes somewhat narrower for higher ice mass fractions because the larger planetary radius allows it to empty out the feeding zone more quickly.

Several interesting effects can be seen.  In the seventh panel showing the ice to rock ratio as a function of time, {\textit{all}} models exhibit the same pattern: initially this ratio is equal to that of the accreted material, but around the time of maximum $\dot{M}$, as the accretion heating causes ice to evaporate, the ratio is lowered.  After $\dot{M}$ decreases again, the loss of water decreases as well, but since a large fraction of the planet has been accreted by this time, the ice to rock ratio remains visibly below its original value.  The amount of mass lost by evaporation is roughly  {up to} 10\% of the final mass of the planet. The peak in $\dot{M}$ corresponds to the peak in the surface temperature (dashed curves in the last panel).
 
\begin{figure*}
	\centering
	\includegraphics[width=0.9\textwidth, trim={0.0cm 1.4cm 0.0cm 0.0cm },clip]{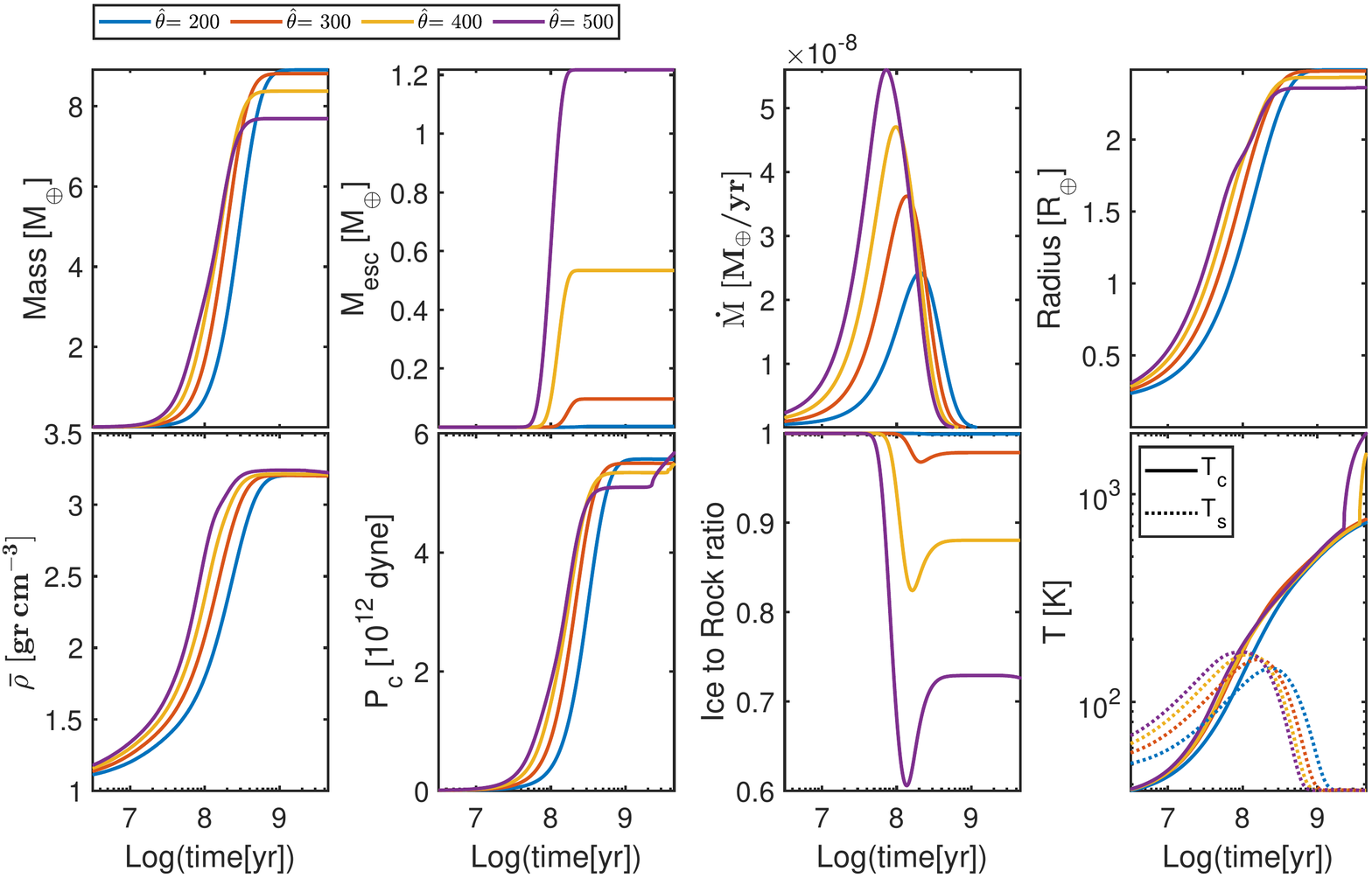}
		\caption{Various planetary properties versus time - as in Fig.\ref{fig:varIce} for different values of $\hat{\theta}$. The models assume an ice mass fraction of 0.5. }
	\label{fig:varS}
\end{figure*}

Another effect can be seen in the plots of the central pressure $P_c$ and central temperature (solid curves in the last panel).  There is a sharp change of slope in the curves for the lower two values of the ice mass fraction.  This is due to radioactive heating after core formation, as mentioned in the previous section. The effect is not seen in models of high ice content. A high ice content has two consequences, which combined, prevent the formation of a rocky (ice-depleted) core. First, the heat capacity of ice is higher than that of rock and secondly, the relatively low content of rock means lower abundances of radioactive species and hence less radioactive heating. As a result, the internal temperatures do not rise sufficiently for the ice to melt and for rock to separate from the ice. The core formation causes a small decrease in the mean density of the body, and this is also visible in the plot of bulk density as a function of time.

In Figure~\ref{fig:varS} we show results for models where the ice mass fraction is kept at $X_{i,b}$, while $\hat\theta$  gradually changes from 200 to 500.  As can be seen from Equation \ref{eq:saf2}, $\dot{M}$ will change by approximately the same factor. In addition, when $\dot{M}$ is larger, the available mass is accreted more efficiently, and the accretion heating is more concentrated in time.  At the highest $\hat\theta$, nearly 20\% of the total mass is lost to water evaporation and  as a result, the final ice to rock ratio is just over 70\% of its original value. For the lowest value of $\hat\theta$, the effect is barely noticeable. 


Only the model corresponding to the highest $\hat\theta$ value forms a large core, and the core starts forming relatively late. We have seen that bodies with ice mass fractions lower than 0.5 are more likely to form cores. In the present case, ice loss during accretion lowered the ice content sufficiently for a core to form.
 
\subsubsection{Final structure - after 4.5 Gyr}

\begin{figure}[h]
	\centering
	\includegraphics[width=1\columnwidth, trim={2.3cm 5.0cm 3.3cm 3.9cm },clip]{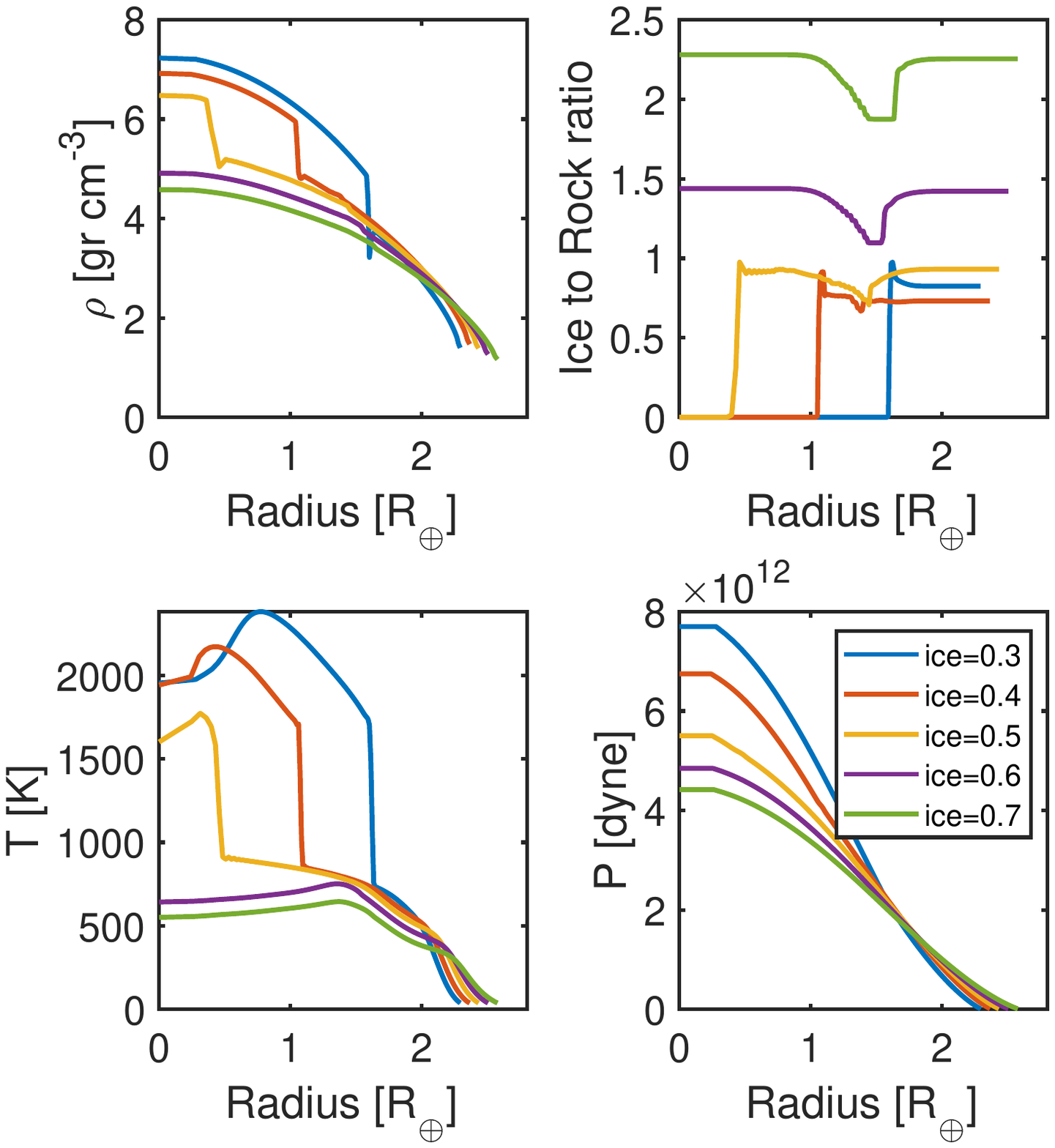}
		\caption{Various planetary properties versus radius in the final stage of planetary evolution, for different compositions. The models assume $\hat{\theta}= 400$. }
	\label{fig:VarIceLast}
\end{figure}

Figure~\ref{fig:VarIceLast} shows the planetary internal structure after $4.5\times 10^9$ years of accretion and evolution.  Profiles are shown for $\hat{\theta}_b$ and different ice mass fractions. For ice mass fractions of 0.5 and above, the density is  smoothly decreasing as a function of radius.  For lower ice mass fractions, where there is sufficient radioactive heating from the rock, so that melting occurs and some of the rock settles towards the center to form a core, the ice to rock ratio is zero in the central part. The extent of the rocky core is larger for the lower ice content. In bodies that do not form a core, one can see in the ice to rock profile a depression corresponding to the accretion phase where ice loss was strongest. The central pressure is inversely proportional to the ice mass fraction, since rock is more compressible than ice.

\begin{figure}[h]
	\centering
	\includegraphics[width=1\columnwidth, trim={2.3cm 5.0cm 3.3cm 3.9cm },clip]{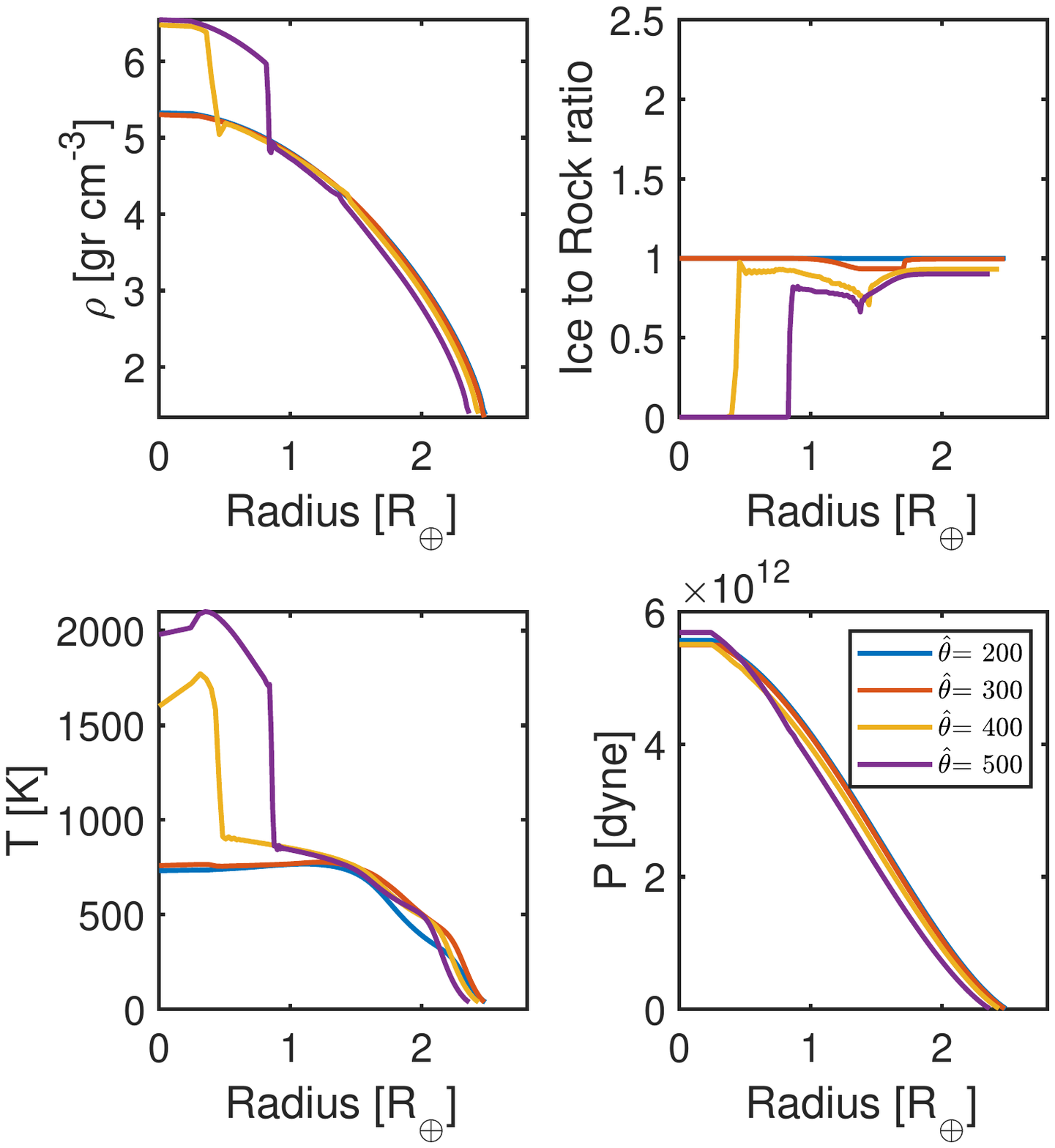}
		\caption{Various planetary properties versus radius in the final stage of planetary evolution, for different normalized Safronov parameters. The models assume 0.5 ice.}
	\label{fig:varSlast}
\end{figure}

Figure~\ref{fig:varSlast} shows how the internal structures vary for different values of $\hat\theta$ for the initial ice mass fraction $X_{i,b}$. For the lower values of $\hat\theta$, the internal structure is quite similar. The effect of lower peak accretion rate, resulting in weaker accretional heating and less sublimation, is seen in the ice to rock profiles. Recall that the 0.5 ice case does not have enough radioactive heating to form a distinct core for $\hat\theta< 400$.  But for $\hat\theta=500$, and $\hat\theta=400$ we clearly see the distinct density discontinuity at near $0.8R_{\oplus}$ and $0.4R_{\oplus}$, respectively, marking the core-mantle boundary.  This is also reflected in the ice to rock ratio and the temperature profile.

\subsubsection{Combined effect of parameters}

We now present the full picture, resulting from the combined effect of the accretion rate---through the Safronov parameter---and the composition---through the initial ice to rock ratio, represented by a set of models corresponding to three values of the normalized Safronov parameter ($\hat\theta=200,400,500$) and three values of the initial ice mass fraction (0.3, 0.5, 0.7) and all 9 combinations between them.

\begin{figure*}
	\centering
	\includegraphics[width=0.9\textwidth, trim={0.0cm 1.4cm 0.0cm 0.0cm },clip]{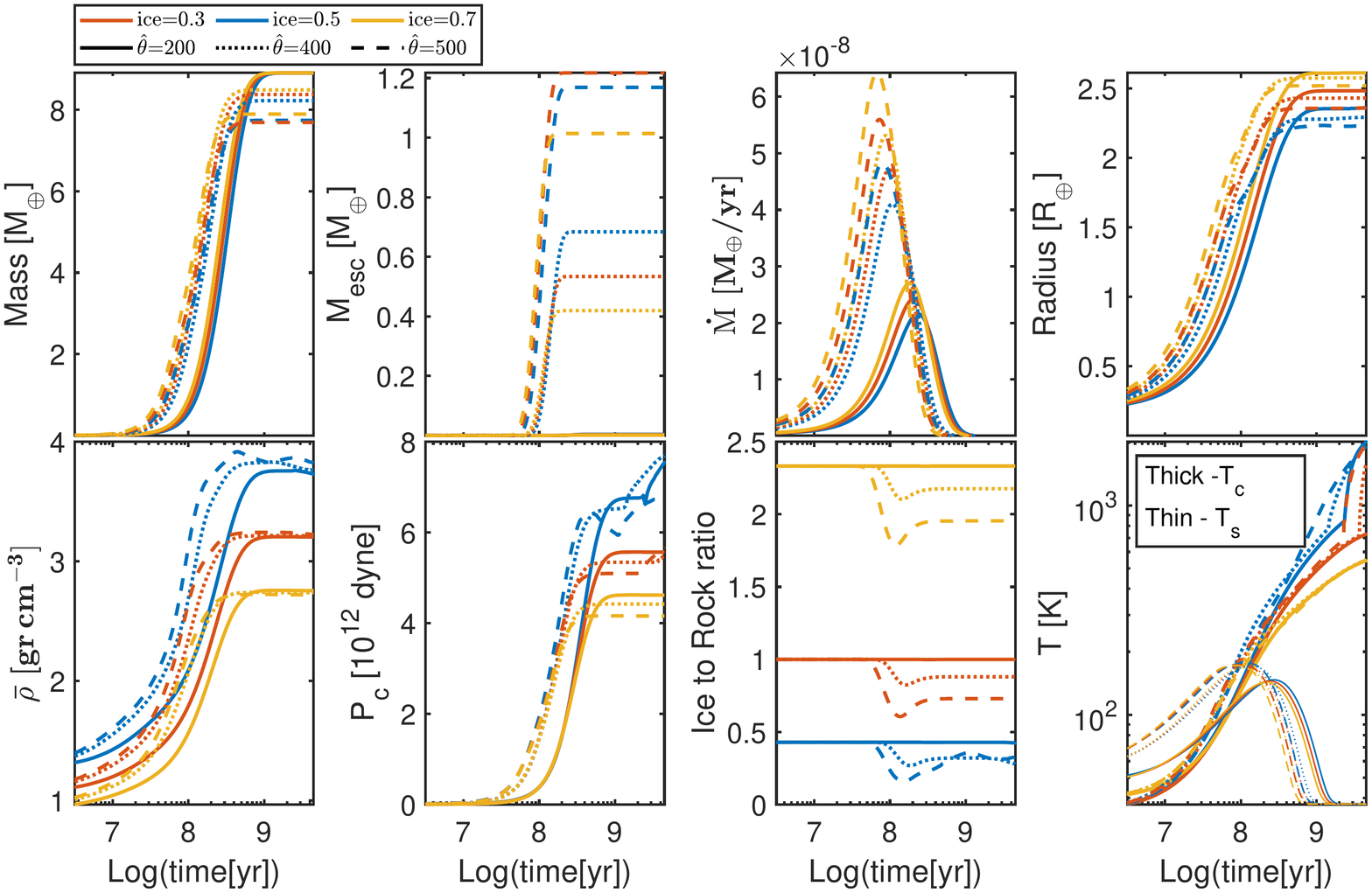}
		\caption{Various planetary properties versus time during planetary evolution, for different compositions and Safronov parameters. }
	\label{fig:Evolution3}
\end{figure*}

\begin{figure}
	\centering
	\includegraphics[width=1\columnwidth, trim={2.3cm 5.0cm 3.3cm 3.9cm },clip]{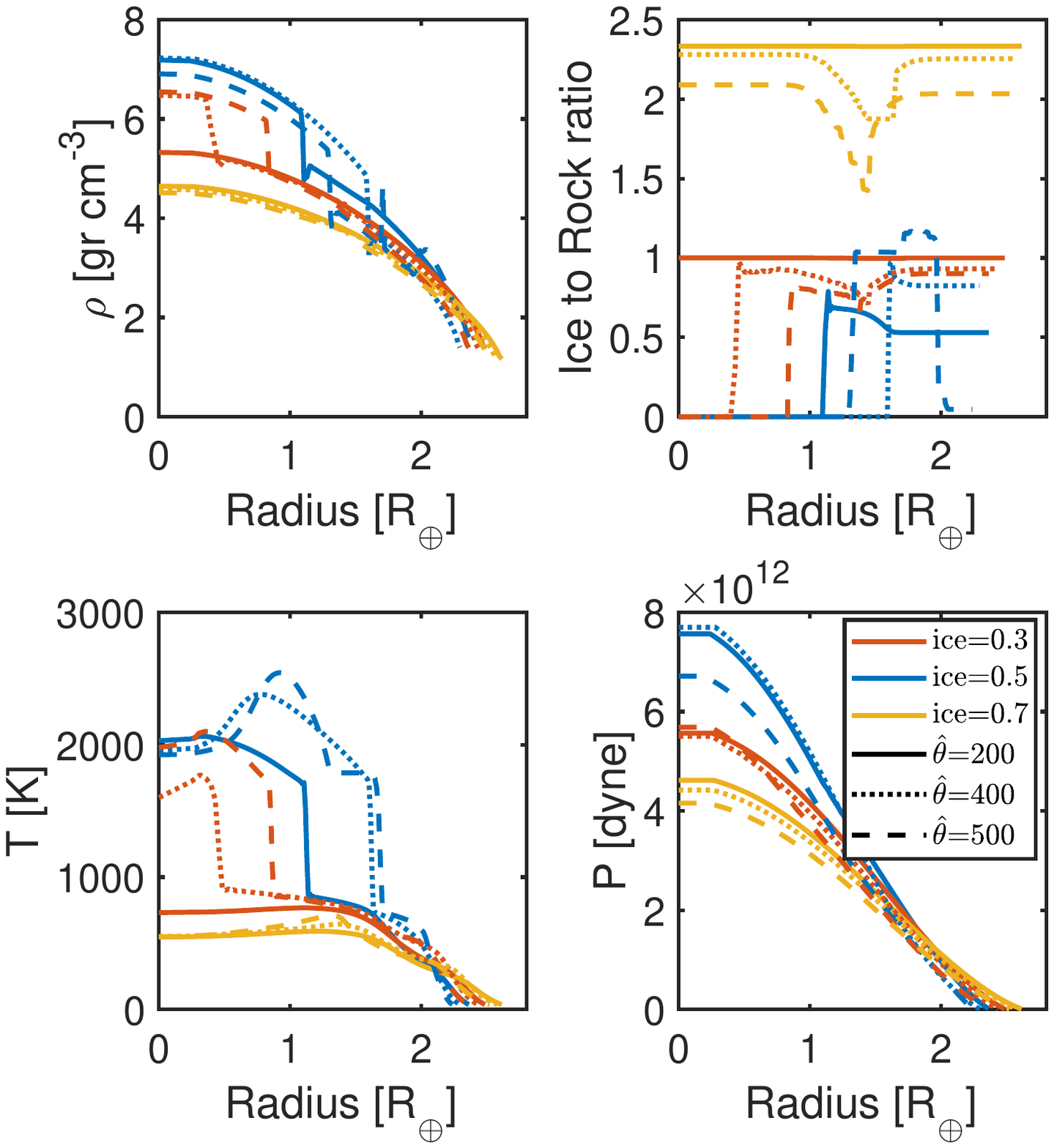}
		\caption{Various planetary properties versus radius in the last stage of planetary evolution, for different compositions and $\hat{\theta}$.}
	\label{fig:lastModles}
\end{figure}

Figure~\ref{fig:Evolution3} shows the combined effect of varying the water ice mass fraction and $\hat\theta$ simultaneously, while Figure \ref{fig:lastModles} does the same for the various parameter profiles after 4.5\,Gyr of evolution. The figures show that a high $\dot{M}$ can help initiate the process of core formation if the rock mass fraction is marginally too low.  Another signature of the accretion rate is seen in the dip in the ice to rock ratio around the radius corresponding to the peak in $\dot{M}$.  




\subsection{Effect of fixed parameters}

\subsubsection{Results for a smaller planetary mass}

In all of the cases presented, the final mass of the planet was almost the same. This is because that mass is essentially determined by the solid mass available in the disk via Equation~\ref{eq:miso}.  In order to gauge the effect of $\dot{M}$ on a smaller mass body, we ran our model for the case of $\Sigma_s=0.1$\,\gcms.  This has the effect of reducing the final mass of the planet by half.

\begin{figure*}
	\centering
	\includegraphics[width=0.9\textwidth, trim={0.0cm 1.4cm 0.0cm 0.0cm },clip]{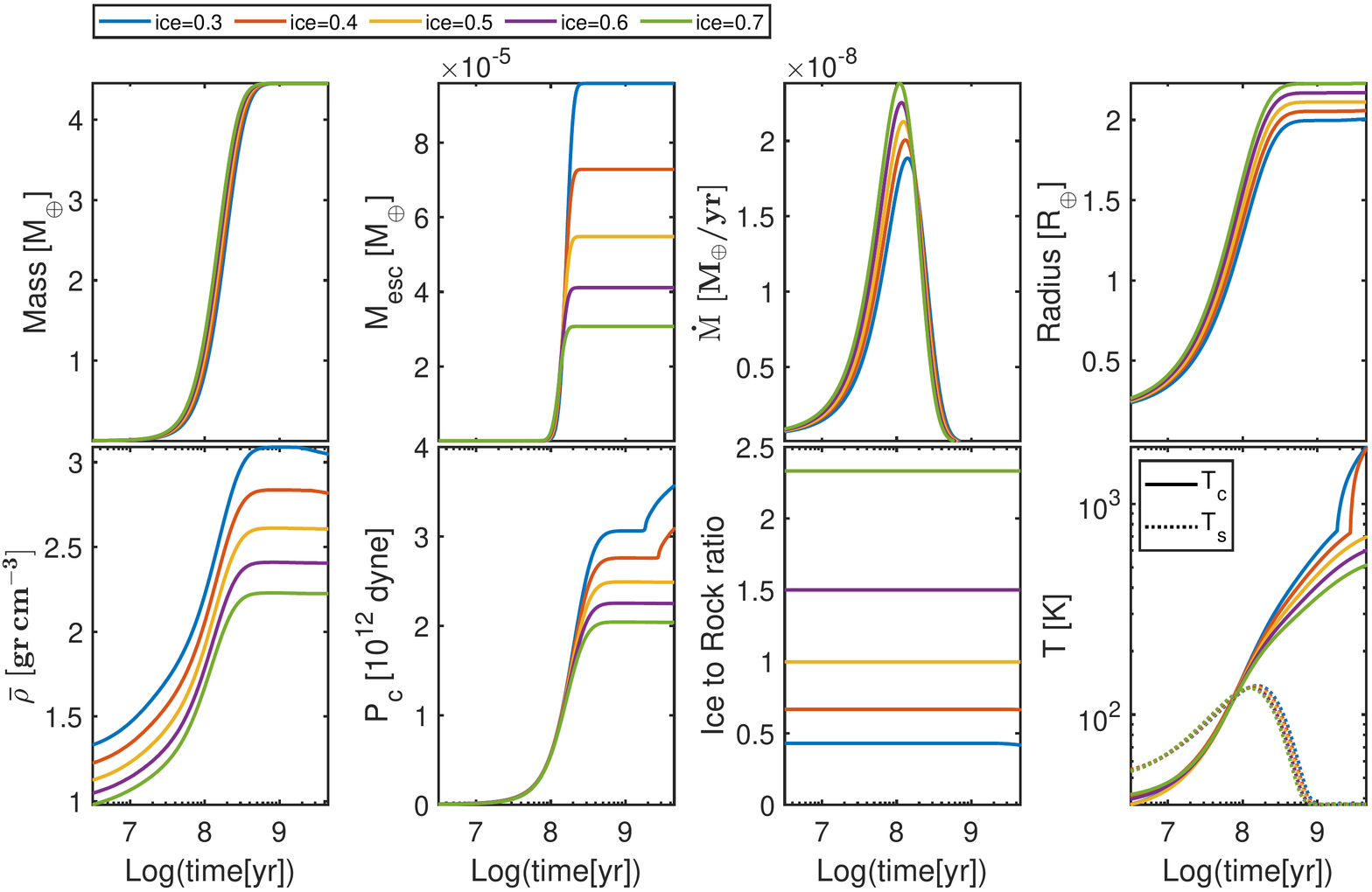}
		\caption{Various planetary properties versus time during planetary evolution, for different ice fractions. The isolation mass is half of original. Can be compared with Figure~\ref{fig:varIce}.}
	\label{fig:EvolutionHalf}
\end{figure*}

\begin{figure}[h]
	\centering
	\includegraphics[width=1\columnwidth, trim={2.3cm 5.0cm 3.3cm 3.9cm },clip]{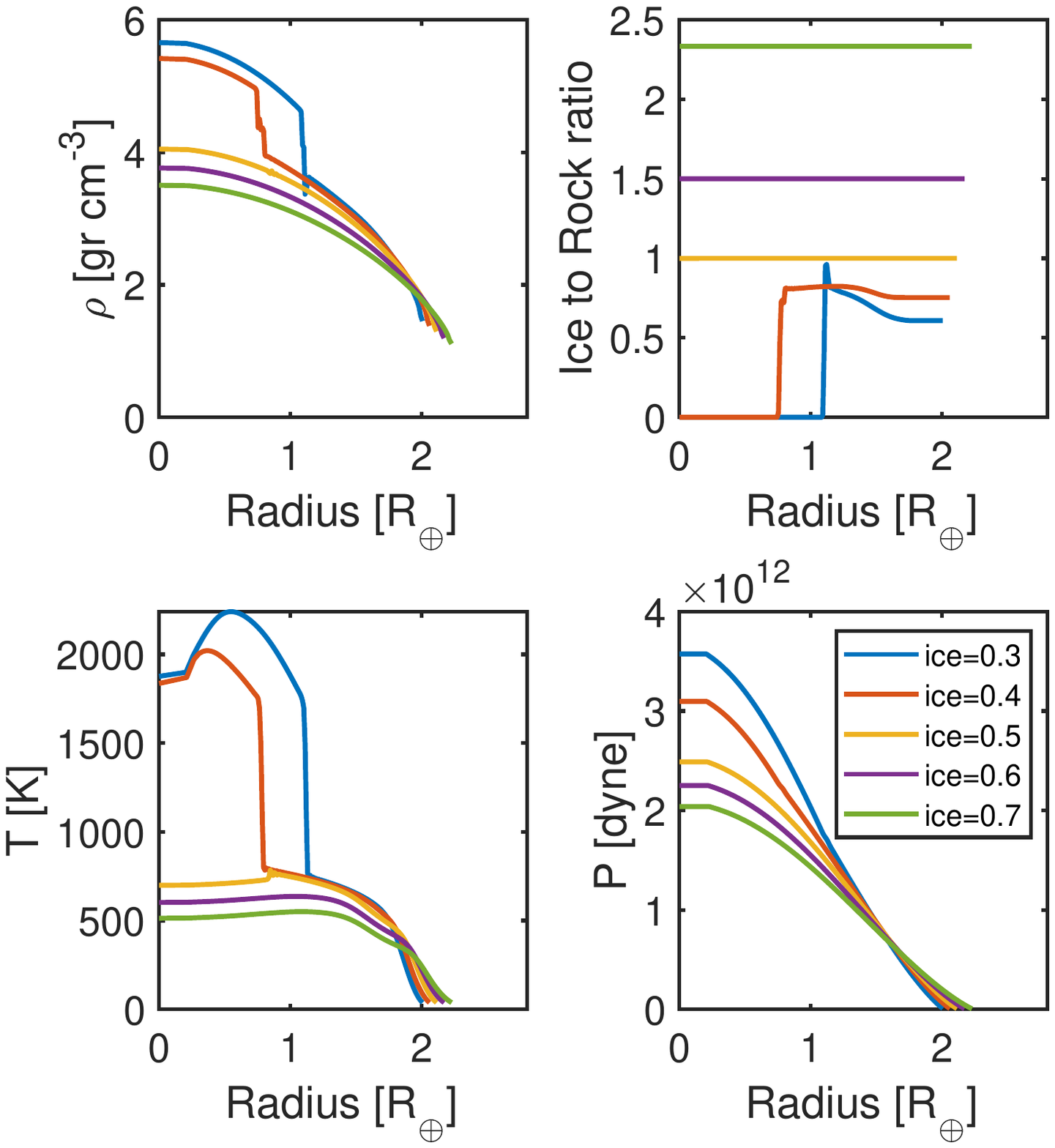}
		\caption{Various planetary properties versus radius in the last stage of planetary evolution, for different compositions, for $\hat{\theta}= 400$. The isolation mass is half of original.}
	\label{fig:lastModlesHalf}
\end{figure}

The results are shown in Figures~\ref{fig:EvolutionHalf} and \ref{fig:lastModlesHalf}. We note that although for these lower-mass models, the densities and pressures are much lower than those of the more massive ones, the temperature profiles are similar, because the radiogenic heat source is the same per unit mass. Similarly, the occurrence and extent of the rocky core are the same. The marked difference is that surface heating by accretion is not sufficient for ice loss, and thus the ice to rock ratio does not differ from the initial value.

\subsubsection{Accretional heating}
\label{accretion}

It is often assumed that accretional heating should raise the interior temperatures of planets to very high values \citep{zeng2014,Boujibar2020}, higher than those reached by our models.
 This implies that the accretion energy (or most of it) has been absorbed by the object during formation. We argue that this assumption is not realistic. In order for the accretion energy, which is essentially a surface source, to be turned into internal  thermal energy, the thermal diffusion timescale $R/\kappa$, where $\kappa$ is the diffusivity, should be comparable to the accretion timescale $M/\dot M$. 
But, for the accretion rate used in Section~\ref{ssec:typical}, corresponding to $\hat\theta=400$, the diffusion timescale is over four orders of magnitude larger than the accretion timescale, and the discrepancy increases for more rapid accretion. 

The question then is, where does the accretion energy go? Some of it goes into ice sublimation (if there is ice in the accreted material), but there is not enough ice to absorb all of it. Most of the energy is simply reradiated. The surface temperature adjusts itself to radiate it, emitting {\textit{accretion luminosity}}. If the rate of energy release by accretion is $\alpha GM\dot M/R$, where $\alpha$ is of order unity, we have
\begin{equation}
 \alpha\frac{GM\dot M}{R}=4\pi R^2 \sigma T_s^4, 
 \label{eq:ts}
\end{equation}
which yields
\begin{equation}
\label{eq:surf}
    T_s=\left(\frac{\alpha G\bar\rho}{3\sigma}\right)^{1/4}\dot M^{1/4}.
\end{equation}
Thus, $T_s$ only needs to be of the order of a few hundred Kelvins in order to  accomplish this. 
If there is ice close to the surface it is controlled by the ice, in which case $\sigma T_s^4$ on the right-hand side of Equation~\ref{eq:ts} is replaced by $Z(T_s)H_v$.
Our detailed results, shown in Figure~\ref{fig:evol}, confirm this estimate.

Large gas planets that form by collapse are different: there, the energy source is a body source that heats the interior. Even before collapse, if the core grows quickly enough, it can attract a hydrogen-helium envelope from the surrounding gas disk before that disk dissipates \citep[e.g.][]{Pollack1996}. In that case the heating comes from the accretional energy of infalling planetesimals and that energy will be deposited inside a massive envelope. Any water vaporization at the core surface due to this accretional energy will be mixed with the H-He envelope and provide heat throughout a large volume, not just at the surface.

Changing the value of $f$---the division of accretion energy between surface and interior---has the expected effect, but does not change our conclusions: a higher $f$ results in more ice loss upon accretion, and converges to a case of an initially lower ice to rock ratio, while a lower $f$, which leaves more ice in the planet, results in lower internal temperatures, similar to the case of an initially higher ice to rock ratio. In both cases, the surface temperature changes only slightly. 

\section{Discussion and Conclusions}
\label{discussion}

The formation of a planet, assuming that it is a continuous process, rather than a cataclysmic event (or a chain of such events), is determined by two major parameters: the accretion rate and the composition of the accreted material. These are the parameters considered in the present study, assuming the composition to be a simple mixture of water ice and silicate rock, and considering continuous spherical accretion. 
Beside understanding evolutionary processes of such planets, our goal is to link the interior structure to the planet's birth environment, as well as to constrain the possible interior structure based on measurements of mass and radius - by no means a straightforward task \citep[e.g.][]{Valencia2006,Rogers2010,Dorn2015}, because different formation scenarios can lead to similar values for the same observables \citep[e.g.][]{Mordasini2012b,Lozovsky2018}.

For this purpose, we have run a large number of models for different combinations of the two leading parameters. The accretion rate is determined by the initial protoplanetary disc structure, in particular its density. Since in any scenario, the accretion rate is proportional to the local density of material and to the growing planetary embryo, the former decreasing with time and the latter, increasing, the rate is time dependent and has a maximum. We find that the functional shape of $\dot M(t)$ has a crucial effect on the final structure of the planet.

It is important to note that different assumptions regarding the mechanism of formation and accretion (such as pebble accretion) lead to very different $\dot{M}$ profiles \citep[e.g.][]{Ida2016,Lambrechts2012,Bitsch2015,Schneider2021,Alibert2018,Grishin2020}. These studies assume that the growth is rapid enough so that the core can attract a significant H-He envelope from the surrounding nebula. It was found, using the pebble accretion paradigm, that sub-Neptune sized planet formation is preferable for relatively low solid accretion rates ($\sim 10^{-6} M_{\oplus}$ yr$^{-1}$ ), which corresponds to low-metallicity environments \citep{Venturini2017}. However, some claim that  pebble accretion plays little or no role at all in terrestrial planet formation \citep{Mah2021} or in planets formed more than a few au from the central star \citep{voelkel2021,voelkel2021a,voelkel2022}. 

We also note that migration during the formation process, which we have not taken into account, can also affect the accretion rate, as well as the composition of accreted material \citep{Alibert2005,Shibata2020,Lega2021,Bitsch2019}.  In the scenario we investigate, $\dot{M}$ is small enough so that the nebular gas dissipates well before a significant core can grow and therefore migration is not important in our case.

{ {Studies of young stellar clusters  indicate that the protostellar gaseous disk dissipates after $\sim$10~Myr \citep{Haisch2001}.  At this time, for the accretion rates we assume, the protoplanetary embryo has a mass  less than $0.05 M_{\oplus}$. At 40 au such a body will have a Bondi radius of $\sim 2\times 10^{10}$~cm.  For the nebular model we are considering, a sphere of this radius would contain $2.6\times 10^{18}\ {\rm g} = 4.4\times 10^{-10}M_{\oplus}$ of nebular gas.  Even if all this gas was accreted onto the embryo, it would comprise a minuscule fraction of the total planet, and therefore we have neglected H-He accretion.}}


 
An important effect of the accretion rate is due to accretion heating, which is proportional to it. We have shown that if heating is intense enough to raise the surface temperature to the sublimation temperature of water ice, then a fraction of the accreted ice escapes ($M_{\rm esc}$). This effect was also found by \cite{Bierson2020}. Therefore the ice to rock ratio of a planet does not necessarily reflect the composition of the environment where it was formed. 
The fate of the escaping vapor into the cold environment is not clear and is not included in our planetary model. While some of it may condense on dust grains, or else drift to other regions of the disc,  some may form a temporary water vapor atmosphere.

{ {It is important to note that the water vapor escapes while the planet grows and therefore it is in constant interaction with the infalling planetesimals. Several studies have shown that planetesimal impacts lead to atmospheric mass loss \citep{Schlichting2015, Biersteker2021, Denman2020, Denman2022, Wyatt2020}. Significant loss may be caused by large impactors \citep{Denman2022}, but the cumulative effect of a large number of small and particularly medium-size impactors may lead to the loss of the atmosphere as well \citep{Schlichting2015}. Assuming a discrete power law size distribution of impacting planetesimals (radii $a_i$)
 with the commonly assumed power index $\gamma=-3.5$ \citep{Wyatt2020}, a mass range of 1~km to a few 1000~km and a total mass equal to the planetary mass $M$, the number of impactors with radius $a_i$ is
 \begin{equation}
     N_i=\frac{3M}{4\pi \rho_p}\frac{a_i^{-\gamma}}{\sum_i a_i^{3-\gamma}},
 \end{equation}
where $\rho_p$ is the density of a planetesimal. The total number of impacts is of the order of $10^{12}$, the number of large impactors ($a_i>1000$~km) is on the order of 10. Extending the sizes distribution to lower radii reduces this number, while a shallower size distribution, say, $\gamma=-2.5$, increases the number of large impactors by more than a factor of 10. We note that the energy required for the vapor to escape is accounted for in the calculation, since only a fraction of the nominal accretion energy is assumed to be absorbed by the growing planet (see Section~\ref{sec:ThermalEvol}), thus in this respect our assumptions are self-consistent. 
A thorough calculation of the effect of impacting planetesimals is beyond the scope of the present study and will be considered in future work, but it seems reasonable to assume that the atmosphere or most of it should be stripped away. We have also tested other processes that can dissipate the atmosphere, such as Jeans escape, hydrodynamic escape or photoevaporation, but found their effect negligible at this distance from the central star.}} In any case, a detailed atmospheric model is necessary to follow the temperature profile and to compute the structure and thermal evolution of such a temporary atmosphere, and to estimate its escape rate. 


Although we have not considered additional volatiles in our study, they will be influenced by the same physics as the water.  \cite{oberg2019} and \cite{bosman2019} have suggested that the uniform enrichment of volatiles in Jupiter's envelope could be due to formation at the outer edge of the solar system where such volatiles could be accreted as ices.  This scenario needs to be revisited taking surface heating into account to see if these ices are indeed retained.

We find that the ice to rock ratio in the environment where the planet grows, determines whether or not a rocky core will form, by rock settling and separation from the ice. This is induced by radiogenic heating by the common long-lived radioisotopes assumed to be embedded in the rock. If the rock content is less than 0.5, the temperature  does not rise sufficiently for the ice to melt and allow rock settling, keeping in mind that the melting temperatures are above $10^3$~K at the internal pressures of the planet. The result is an almost homogeneous configuration. Even when a core forms, the displaced ice increases the ice to rock ratio in the mantle, but there is no pure ice layer. Between the rocky core and the ice-rich mantle, a relatively thin layer is found, where superheated water is mixed with the rock. The temperature is however too high for the water to be chemically absorbed by the rock.

\begin{figure}[h]
\centering
\includegraphics[width=0.475\columnwidth]{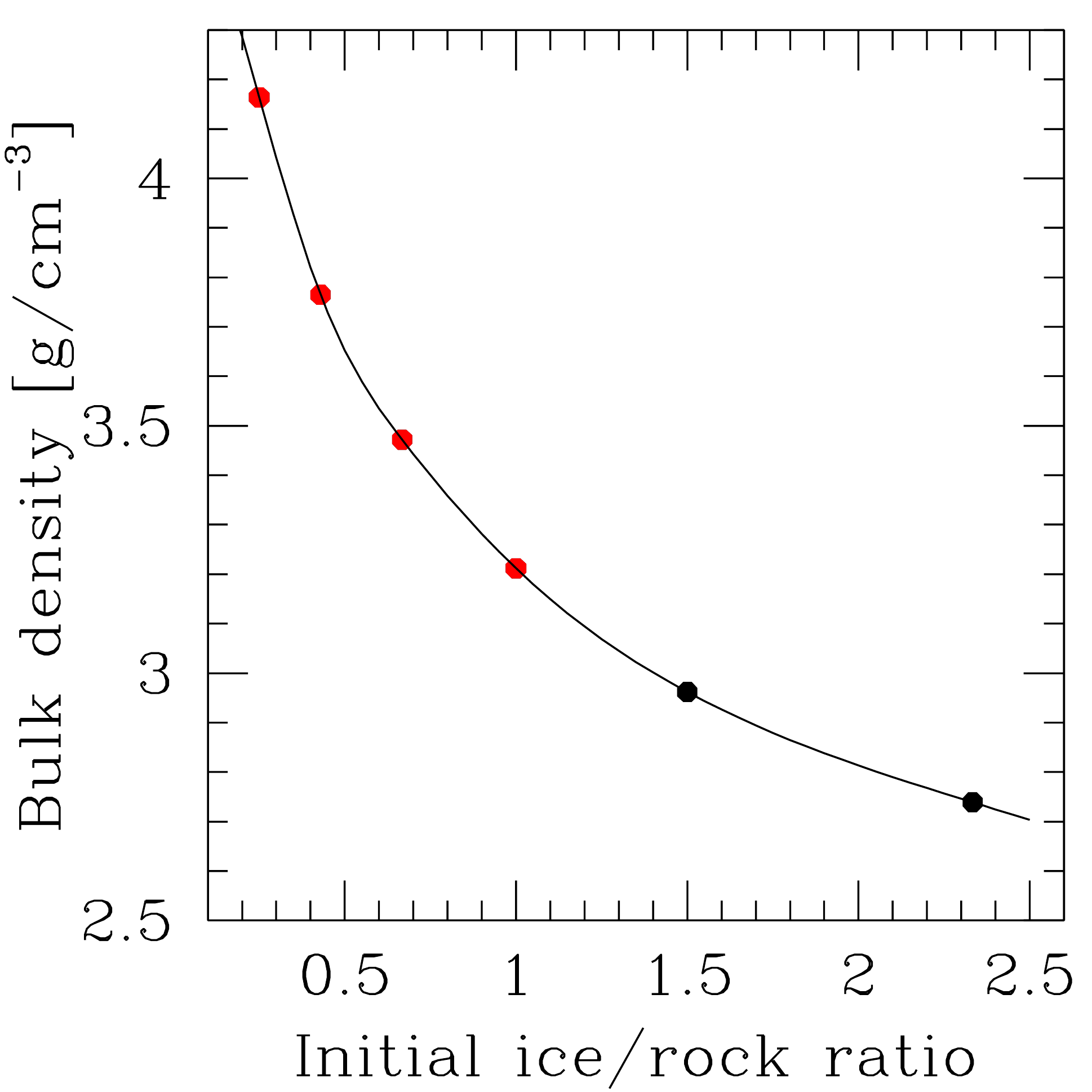}
\includegraphics[width=0.475\columnwidth]{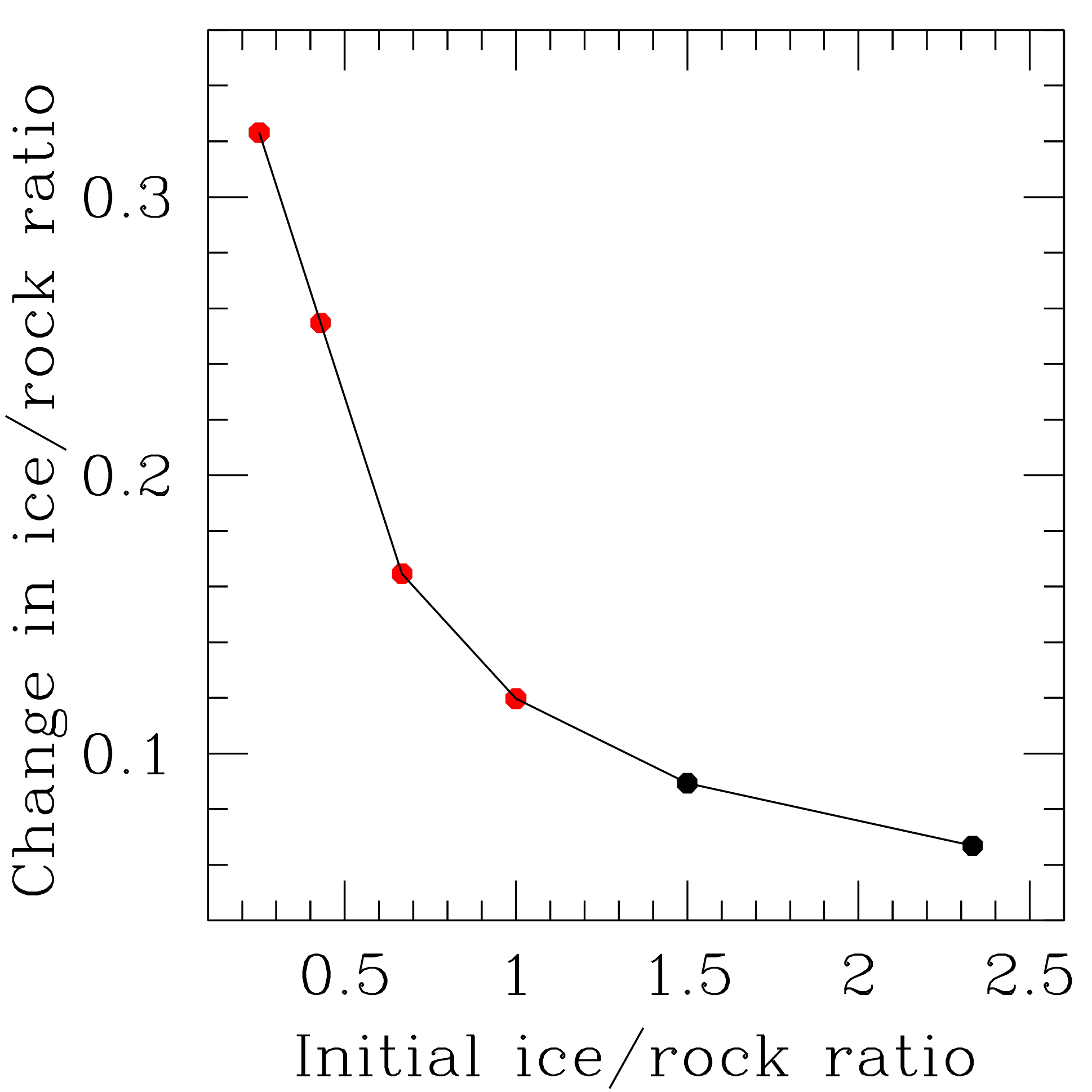}
\caption{{\textit{Left:}} Bulk density of the final structure of models with $\hat\theta=400$ as a function of the initial ice to rock ratio. Red points mark models that have developed a rocky core. {\textit{Right:}} Relative change of the ice to rock ratio due to loss of ice during accretion, as a function of the initial ratio.}
\label{fig:ratio}
\end{figure}

In principle, if the mass of a planet and its radius are known, so that the bulk density can be derived, models like those presented here may be used to derive the planet's ice to rock ratio and to determine whether the planet possesses a rocky core. Furthermore, the same models will reveal what was the ice to rock ratio in the environment where the planet formed. This is illustrated in Figure~\ref{fig:ratio}, which shows---for a given Safronov parameter---the dependence of bulk density and the relative change in ice to rock ratio as a function of initial ice to rock ratio.

\begin{figure}[h]
	\centering
	\includegraphics[width=1\columnwidth, trim={3.1cm 6.2cm 3.6cm 5.2cm },clip]{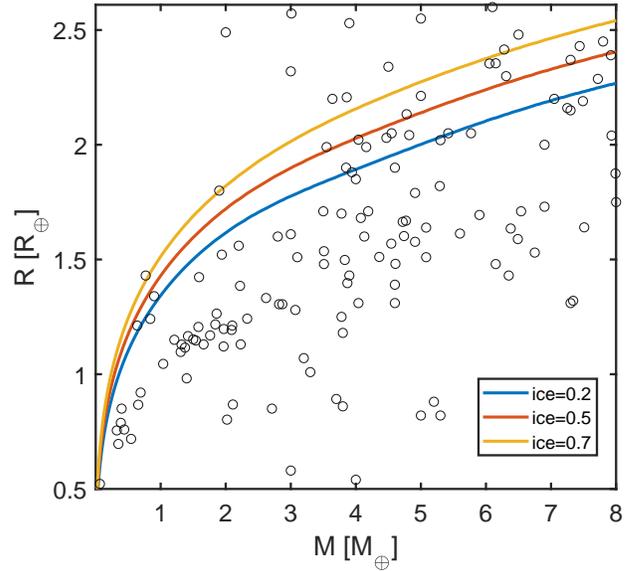}
		\caption{ Mass - Radius relation for models with various ice mass fractions and $\hat\theta =400$. Circles denote the sample of discovered exoplanets, up to 8 $M_\oplus$ (based on \cite{NASA_Exoplanet}). See text for discussion.}
	\label{fig:M-R-exo}
\end{figure}

{ {Our results may be seen in the light of exoplanets.  In Figure~\ref{fig:M-R-exo} we show the $R-M$ relation for the baseline model on the background of observed exoplanets. While the scatter of exoplanets in the $R-M$ plane is considerable, there are numerous objects in the region of our R(M) relation, indicating an ice-rich composition (in the absence of a H-atmosphere). Admittedly, the observed exoplanets are much closer to the host star than our models, but if these bodies were formed in debris disks after the nebular gas has dissipated, the accretion mechanism could be similar to the one we consider. In addition, the orbital distance of 40 au used in our study was coupled with the assumption of a solar-like central star; the same boundary condition may be obtained much closer to a fainter star.  Moreover, the stellar radiation contributes only a small fraction of the energy for a protoplanet, as shown in Figure~\ref{fig:evol}, and therefore, the formation distance may be smaller still. There is however a lower limit, imposed by the snow line (demanding that the local equilibrium temperature be lower than 170-180~K). Planets that are now closer to the host star, but have densities indicating an ice-rich composition, may have migrated inward since formation \citep[][and references therein]{Bean2021}. For now, the snow line is beyond the detection limit of small planets, but we may expect more ice-rich super-earths to be discovered in future.}}  

A very interesting study of the Trappist-1 system by \cite{coleman2019} investigated how the ice content of these planets differed for planets formed via pebble accretion as compared to planets formed via planetesimal accretion.  This study should be revisited taking the physics of evaporation at the surface into account.  More careful treatment of the details of the accretion process can shed new light on the modeling of exoplanet evolution.

Finally, one of the interesting results of this work is the dependence of the final composition on the Safronov parameter (see e.g.,  Figure~\ref{fig:lastModles})---for relatively high values of $\hat\theta$, the ice mass fraction in the planets is lower than the assumed ice mass fraction of the planetesimals. However, one should keep in mind that the Safronov parameter indicates the efficiency of accretion and might depend on the mass spectrum \citep[e.g.][]{Namouni1996}. Therefore, the ice to rock ratio of the accreted material and the solid surface density of the disk might influence $\hat\theta$. Hence it is possible that the two parameters considered here as independent, may be linked. This would make the present study much more conclusive by eliminating some of the parameter combinations.

\begin{acknowledgments}
\textit{We acknowledge the support of the Israeli Science Foundation for this research through the ISF grant 566/17.}
\end{acknowledgments}

\newpage

\appendix

\section{List of models}
\setcounter{table}{1}
\begin{table}[h]
\begin{tabular}{|llllllll}
\hline
\multicolumn{1}{|l|}{ice mass fraction} & \multicolumn{1}{l|}{$\hat{\theta}$} & \multicolumn{1}{l|}{Fig.\ref{fig:varIce}, \ref{fig:VarIceLast} } & \multicolumn{1}{l|}{Fig. \ref{fig:varS}, \ref{fig:varSlast} } & \multicolumn{1}{l|}{Fig. \ref{fig:Evolution3}, \ref{fig:lastModles}} & \multicolumn{1}{l|}{Fig. \ref{fig:EvolutionHalf}, \ref{fig:lastModlesHalf} } & \multicolumn{1}{l|}{M ($M_{\oplus}$)}      & \multicolumn{1}{l|}{R ($R_{\oplus}$)}      \\ \hline
\multicolumn{8}{|c}{Full Mass}\\ \hline
\multicolumn{1}{|l|}{0.3}                & \multicolumn{1}{l|}{200}         & \multicolumn{1}{l|}{}     & \multicolumn{1}{l|}{}      & \multicolumn{1}{l|}{x}     & \multicolumn{1}{l|}{}      & \multicolumn{1}{l|}{8.29} & \multicolumn{1}{l|}{2.30} \\ \hline
\multicolumn{1}{|l|}{0.5}                & \multicolumn{1}{l|}{200}         & \multicolumn{1}{l|}{}     & \multicolumn{1}{l|}{x}     & \multicolumn{1}{l|}{x}     & \multicolumn{1}{l|}{}      & \multicolumn{1}{l|}{8.90} & \multicolumn{1}{l|}{2.48} \\ \hline
\multicolumn{1}{|l|}{0.7}                & \multicolumn{1}{l|}{200}         & \multicolumn{1}{l|}{}     & \multicolumn{1}{l|}{}      & \multicolumn{1}{l|}{x}     & \multicolumn{1}{l|}{}      & \multicolumn{1}{l|}{8.91} & \multicolumn{1}{l|}{2.61} \\ \hline
\multicolumn{1}{|l|}{0.5}                & \multicolumn{1}{l|}{300}         & \multicolumn{1}{l|}{}     & \multicolumn{1}{l|}{x}     & \multicolumn{1}{l|}{}      & \multicolumn{1}{l|}{}      & \multicolumn{1}{l|}{8.82} & \multicolumn{1}{l|}{2.48} \\ \hline
\multicolumn{1}{|l|}{0.3}                & \multicolumn{1}{l|}{400}         & \multicolumn{1}{l|}{x}    & \multicolumn{1}{l|}{}      & \multicolumn{1}{l|}{x}     & \multicolumn{1}{l|}{}      & \multicolumn{1}{l|}{8.29} & \multicolumn{1}{l|}{2.30} \\ \hline
\multicolumn{1}{|l|}{0.4}                & \multicolumn{1}{l|}{400}         & \multicolumn{1}{l|}{x}    & \multicolumn{1}{l|}{}      & \multicolumn{1}{l|}{}      & \multicolumn{1}{l|}{}      & \multicolumn{1}{l|}{8.32} & \multicolumn{1}{l|}{2.36} \\ \hline
\multicolumn{1}{|l|}{0.5}                & \multicolumn{1}{l|}{400}         & \multicolumn{1}{l|}{x}    & \multicolumn{1}{l|}{x}     & \multicolumn{1}{l|}{x}     & \multicolumn{1}{l|}{}      & \multicolumn{1}{l|}{8.37} & \multicolumn{1}{l|}{2.43} \\ \hline
\multicolumn{1}{|l|}{0.6}                & \multicolumn{1}{l|}{400}         & \multicolumn{1}{l|}{x}    & \multicolumn{1}{l|}{}      & \multicolumn{1}{l|}{}      & \multicolumn{1}{l|}{}      & \multicolumn{1}{l|}{8.43}  & \multicolumn{1}{l|}{2.50} \\ \hline
\multicolumn{1}{|l|}{0.7}                & \multicolumn{1}{l|}{400}         & \multicolumn{1}{l|}{x}    & \multicolumn{1}{l|}{}      & \multicolumn{1}{l|}{x}     & \multicolumn{1}{l|}{}      & \multicolumn{1}{l|}{8.49} & \multicolumn{1}{l|}{2.58} \\ \hline
\multicolumn{1}{|l|}{0.3}                & \multicolumn{1}{l|}{500}         & \multicolumn{1}{l|}{}     & \multicolumn{1}{l|}{}      & \multicolumn{1}{l|}{x}     & \multicolumn{1}{l|}{}      & \multicolumn{1}{l|}{7.82} & \multicolumn{1}{l|}{2.24} \\ \hline
\multicolumn{1}{|l|}{0.5}                & \multicolumn{1}{l|}{500}         & \multicolumn{1}{l|}{}     & \multicolumn{1}{l|}{x}     & \multicolumn{1}{l|}{x}     & \multicolumn{1}{l|}{}      & \multicolumn{1}{l|}{7.69} & \multicolumn{1}{l|}{2.36} \\ \hline
\multicolumn{1}{|l|}{0.7}                & \multicolumn{1}{l|}{500}         & \multicolumn{1}{l|}{}     & \multicolumn{1}{l|}{}      & \multicolumn{1}{l|}{x}     & \multicolumn{1}{l|}{}      & \multicolumn{1}{l|}{7.92} & \multicolumn{1}{l|}{2.52} \\ \hline
\multicolumn{8}{|c|}{Half Mass}                                                                                                                                                                                                                           \\ \hline
\multicolumn{1}{|l|}{0.3}                & \multicolumn{1}{l|}{400}         & \multicolumn{1}{l|}{}     & \multicolumn{1}{l|}{}      & \multicolumn{1}{l|}{}      & \multicolumn{1}{l|}{x}     & \multicolumn{1}{l|}{4.45} & \multicolumn{1}{l|}{2.01} \\ \hline
\multicolumn{1}{|l|}{0.4}                & \multicolumn{1}{l|}{400}         & \multicolumn{1}{l|}{}     & \multicolumn{1}{l|}{}      & \multicolumn{1}{l|}{}      & \multicolumn{1}{l|}{x}     & \multicolumn{1}{l|}{4.45} & \multicolumn{1}{l|}{2.06} \\ \hline
\multicolumn{1}{|l|}{0.5}                & \multicolumn{1}{l|}{400}         & \multicolumn{1}{l|}{}     & \multicolumn{1}{l|}{}      & \multicolumn{1}{l|}{}      & \multicolumn{1}{l|}{x}     & \multicolumn{1}{l|}{4.45} & \multicolumn{1}{l|}{2.11}  \\ \hline
\multicolumn{1}{|l|}{0.6}                & \multicolumn{1}{l|}{400}         & \multicolumn{1}{l|}{}     & \multicolumn{1}{l|}{}      & \multicolumn{1}{l|}{}      & \multicolumn{1}{l|}{x}     & \multicolumn{1}{l|}{4.45} & \multicolumn{1}{l|}{2.17} \\ \hline
\multicolumn{1}{|l|}{0.7}                & \multicolumn{1}{l|}{400}         & \multicolumn{1}{l|}{}     & \multicolumn{1}{l|}{}      & \multicolumn{1}{l|}{}      & \multicolumn{1}{l|}{x}     & \multicolumn{1}{l|}{4.45} & \multicolumn{1}{l|}{2.23} \\ \hline
\end{tabular}\caption{List of models used in this study and figures there they presented. M and R are the final mass and radius of the bodies. "Half Mass" is referring to models with a isolation mass set to be half of the original ("Full Mass"). }\label{tab:modelList}
\end{table}

In Table \ref{tab:modelList}.A we list the models used in this study, with the corresponding figures numbers.

\newpage
\bibliography{Main}{}
\bibliographystyle{aasjournal}

\end{document}